\documentclass[runningheads,a4paper]{llncs}

\usepackage[colorlinks=true,linkcolor=blue,anchorcolor=black,citecolor=magenta,filecolor=black,menucolor=black,runcolor=black,urlcolor=black]{hyperref}

\usepackage{natbib}
\usepackage[bottom]{footmisc}

\usepackage{color,soul}
\setulcolor{red}
\sethlcolor{yellow}

\usepackage{mathtools}
 
\usepackage{array}
\newcolumntype{P}[1]{>{\centering\arraybackslash}p{#1}}

\usepackage{enumitem}
\newlist{tabitem}{itemize}{1}
\setlist[tabitem]{wide=0pt, nosep, leftmargin= * ,label=\textbullet,after=\vspace{-\baselineskip},before=\vspace{-0.6\baselineskip}}

\usepackage[utf8]{inputenc}
\usepackage{amsmath}

\usepackage{geometry}
\usepackage{array}

\usepackage{amssymb}
\usepackage{graphicx}

\usepackage{url}
\begin{document}

%\mainmatter  % start of an individual contribution

\title{Comparison of Different Configurations of Saturated Core Fault Current Limiters in a Power Grid by Numerical Method}

\titlerunning{Comparison of Different Configurations of Saturated Core}

\author{Aydin Zaboli}


\institute{Department of Electrical Engineering,\\ Amirkabir University of Technology, Tehran, Iran \\
\email{a.zaboli@aut.ac.ir}}


%\toctitle{Lecture Notes in Computer Science}
%\tocauthor{Authors' Instructions}
\maketitle

\begin{abstract}
Short circuit fault currents are increasing due to growing demand for electricity and high complexity in power systems. Because the fault currents reach the highest value which the breakers are unable to restrict, the electrical grid's security is under jeopardy. By entering a limiting impedance into a transmission line in series, these impedances restrict the rising amounts of fault currents to levels that are acceptable. Saturated core fault current limiters (SCFCLs) are a pivotal tool for limiting fault currents rise in power networks that have good performance characteristics. In a normal condition, these limiters have slight effects on the system and can effectively limit short-circuit currents when occur. In this chapter, various structures of SCFCLs with different arrangements of ac windings \& dc windings are presented and the currents passing through the FCLs under the normal and faulty system conditions are assessed and compared. The flux density in various regions of the core in different arrangements has been investigated as well and the desired analyzes have been performed. Simulation will be presented based on COMSOL Multiphysics 5.4, a finite element software package which can provide a precious assessment to compare these protective devices with different configurations.\footnote{This is the author's pre-print of a book chapter accepted for publication in ``Modernization of Electric Power Systems'' Springer.}

%\keywords{We would like to encourage you to list your keywords within the abstract section}
\end{abstract}

\section{Introduction}
Increase in fault currents in power grids is due to the ongoing expansion in energy generation and demand, as well as the trend to parallelize the network sections. As a result, there is a global effort underway to find solutions to the fault currents problem in order to prevent having to replace existing circuit breakers or to postpone their replacements \cite{naphade2021experimental,jia2017numerical,zhao2014performance,zhou2020hybrid,yuan2020optimized,zhou2019performance,liu2021design,wei2020limiting,ouali2020integration,alam2018fault}. Using various types of fault current limiters (FCLs) could be an option. This device essentially makes a series connection to the power grid with a variable impedance. The impedance is low when it performs at its best situation. When a fault arises, however, the impedance of the system increases dramatically, preventing amplitude currents from flowing to greater levels. Since the failures in the converters and rotor circuit might result dangerously in FCLs, on the other hand, may be employed to maintain the generator's grid connection for as long as possible. Short circuit current is limited by such circuits, but the generator is not disconnected from the grid \cite{zhang2019viable,li2019current,shen2018three}. Despite being advantageous to utilities, these new methods have created a significant problem in terms of short-circuits. Symmetrical or asymmetrical short circuits are unavoidable in power networks and result in large magnitude short circuit currents, particularly in high voltage power networks. These currents have negative thermal and mechanical consequences \cite{safaei2020survey}. Different types of these devices can be classified by their functionalities, materials, structures, expenditures, etc. They may be classified into two general groups based on the most significant functional characteristics \cite{shen2018study,wei2018performance,yuan2018novel,baferani2018novel}: 1- Limiters that have a built-in reaction to a fault. 2- Limiters with a delayed error response. SCFCLs offer numerous benefits over other types, however, they necessitate a significant quantity of magnetic substances, resulting in a high upfront cost and a considerable scale. These FCLs can be as non-superconducting and superconducting which our attention is on non-superconducting FCLs in this chapter. Superconducting FCLs have at least a superconductor tape or coil in their configurations which can be effective in limiting the fault currents, however, there are some problems in using these types of limiters. Some issues are such as cooling system volume and superconductor costs that make barriers for researchers and engineers to provide a suitable design of power grid including the superconducting FCLs \cite{jia2017numerical,ruiz2014resistive,tan2015resistive,chen2016parameter,commins2012three}. Nowadays these FCLs are constructed in small scales in laboratories and test in some small sections of power grids \cite{tan2015resistive}. There are different types of FCLs including saturated core FCL, inductive FCL, magnetic FCL, open core FCL, etc. that every configuration has special structure and material, but the same functionality.\\
The saturated core FCL is made up of three parts including copper dc \& ac coils, and an iron core, each of which has its own structure. The power supply is linked in series to the ac coils, while the dc coils are connected to a separate direct current source. These sorts of limiters, on the other hand, confront problems such as increasing the needed magnetic material, which maybe partially addressed by introducing novel structures and arrangements. To address the issue of big volume and high cost, many structures are offered \cite{shen2018three,shen2018study,linden2019design,linden2020phase,commins2012analytical}. By introducing an air gap into the core, the volume of the magnetic substance may be decreased. This approach is applicable to nonlinear inductors in general. The air gap is not just a component of an alternating current circuit, but it is also a component of a direct current magnetic circuit. In this situation, core saturation is difficult, and the FCL's usual impedance is decreased. As a result, designs with three legs are possible, and the arrangement of ac and dc windings may be modified such which there is no dc flux passing through the middle leg of the core. As a result, an air gap in the intermediate branch may be generated without raising the dc excitation to saturate the nucleus. This air gap also guarantees that the core size is reduced for the same voltage level. The configuration of the ac and dc windings allows one of the two outer legs to go out of saturation during each half cycle of the fault, limiting the fault current. The weight of the FCL may be decreased by utilizing the proper construction when compared to induction type designs, which depends on how the ac and dc coils are positioned and put on distinct legs. The ratio of the defective phase voltage to the overall fault current determines the effective fault impedance \cite{cvoric2009new,cvoric20163d,cvoric2008comparison}.

\section{Literature Review}
There are many FCL configurations tested or implemented all over the world to relieve the risks of these hazardous currents in a power grid. These can help the power system to be safe, reliable, and stable \cite{safaei2020survey}. Ruiz et al. \cite{ruiz2014resistive} has been investigated the Resistive-type Superconducting FCLs in terms of their working principles, numerical modeling, and superconducting materials with experimental concepts. Tan et al. presented a resistive SFCL with different current flowing time by an YBCO tape as a superconductor material \cite{tan2015resistive}. Assessment of thermoelectric model of RSFCL by root mean squared method is conducted by Branco et al. in \cite{branco2010proposal}. Related to inductive SFCL, Yamaguchi et al. presented a transformer type SFCL that analyze the relation between the current limiting characteristics and transformer ratings in \cite{yamaguchi2004characteristics}. Recently, a transformer-type SFCL with an external circuit has been developed that may minimize the fault current utilizing twice quench operations. The impact of current limitation on the winding path of two isolated windings has been addressed \cite{han2018fault}. A comprehensive review of Flux-locked type FCLs (FLSFCLs) was recently published in \cite{badakhshan2018flux}, which included a detailed assessment of research conductions and applications of FLSFCLs in power grids. A FLSFCL was used to enrich the quality of power indexes in a distribution system \cite{lim2011analysis}. A voltage-sourced inverter has been used to incorporate the planned FLSFCL. The coordination of overcurrent relays with FLSFCL is discussed in \cite{kim2010study}. In \cite{zhao2014performance}, the impacts of these windings on FLSFCL’s performance in an iron closed-core were investigated. Tripathi et al. \cite{tripathi2021real} has been conducted a configuration with two ac winding and a dc winding SCFCL for improving a doubly fed induction generator (DFIG) system.

\section{Principles of Fault Current Limiting}
FCLs are widely studied during recent years in which researchers are introduced many configurations with different material, performance type, arrangements, etc. Two fault current liming basis are as follows: Adding resistive impedance and reactive impedance (an inductor or a combination of an inductor and a capacitor) \cite{safaei2020survey}.\\
There is a sample transmission line with FCL is represented in which this device is in series with other power grid components. FCLs usually limit fault currents to an acceptable level which it is relied on the system requirements. That means they do not act during normal conditions of a power grid. Also, they do not apply any changes, damage, or restrictions to the system, so could be safe and reliable.\\

\begin{figure}
\centering
\includegraphics[height=1.5cm]{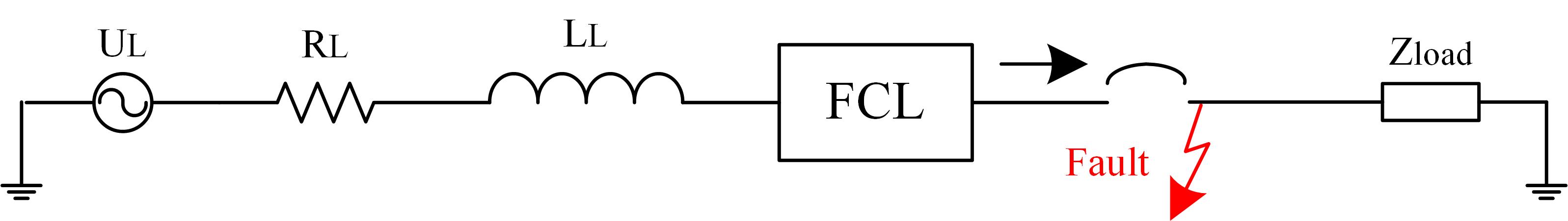}
\caption{Sample power system circuit}
\label{fig:Fig. 2.1}
\end{figure}

Eq. (\ref{1}) may describe the fault current in the system, such as the single-phase fault to ground (see Fig.~\ref{fig:Fig. 2.1}) in which $i_{fault(t)}$ represents the fault current, $U_{L(t)}$, the line voltage, $L$ the phase angle of the transmission line impedance, $Z_{L}$ the impedance of transmission line, $I_{L, t_{f}}$ the value of a current of transmission line at the start time during a fault condition, $t_{f}$ the start time of a fault, and $\tau$ the line time constant \cite{yuan2018novel}:

\begin{equation}\label{1}
    {i_{fault}(t) = \frac{U_{L} \times sin(\omega t - \phi_{L})}{|Z_{L}|} + (I_{L,t_{f}}-\frac{U_{L} \times sin(\omega t - \phi_{L})}{|Z_{L}|}) \times e^{-\frac{t-t_{f}}{\tau}}}
\end{equation}

When a resistive or inductive types of fault current limiters are utilized, the difference in system’s fault performance is in the time constant value of the system, resulting in a difference in the value of the fault current's dc portion. The transmission line time constant of a fault is less than the inductive FCL for the same FCL impedance when a resistive FCL is employed, as demonstrated by Eqs. (\ref{2}) and (\ref{3}) which $L_{L}$ is line leakage inductance, $R_{L}$, line leakage resistance, and $R_{FCL}$ and $L_{FCL}$ are resistive and inductive impedances, correspondingly.

\begin{equation}\label{2}
    {\tau_{RFCL} = \frac{L_{L}}{R_{L} + R_{FCL}}}
\end{equation}

\begin{equation}\label{3}
    {\tau_{XFCL} = \frac{L_{L} + L_{FCL}}{R_{L}}}
\end{equation}

As a result, given the same FCL impedance, the initial fault current peak for resistive limiting impedance is less than the inductive impedance. Another distinction is that the resistive FCL dissipates energy during the fault, whereas the inductive FCL accumulates energy in the magnetic field during the failures and restores it to the system at the end of each cycle. Hence, inductive FCLs do not result in power loss for the system if normal system operation is resumed without interruption of current flow (regardless of inductor’s resistance). When the transmission line is terminated, the energy accumulated in the previous cycle is merely the power loss in the circuit breaker \cite{tan2015resistive,zhou2020inductive}.\\
To limit the fault current, inductive FCL can employ both an inductor and a capacitor, simultaneously. The FCL can be substituted by a parallel inductor and capacitor in this case to arrange to conduct the network frequency to a resonance state. The $C_{FCL}$ capacitor is connected in series via a transmission line, and its quantity is adjusted to compensate the leakage inductance of the transmission line. By occurring a fault, the $L_{FCL}$ inductance is connected in parallel to the capacitor, and the parallel LC circuit's resonant impedance restricts the fault current as follows:

\begin{equation}\label{4}
    {Z_{res,FCL} =j \times \frac{\omega L_{FCL}}{1-\omega^2 L_{FCL}C_{FCL}}}
\end{equation}

This section briefly discussed the limitation of fault currents. Then a comprehensive comparison will be given to assess the different configurations by numerical method in the next sections.

\section{Basis of SCFCLs}
\begin{figure}
\centering
\includegraphics[height=6cm]{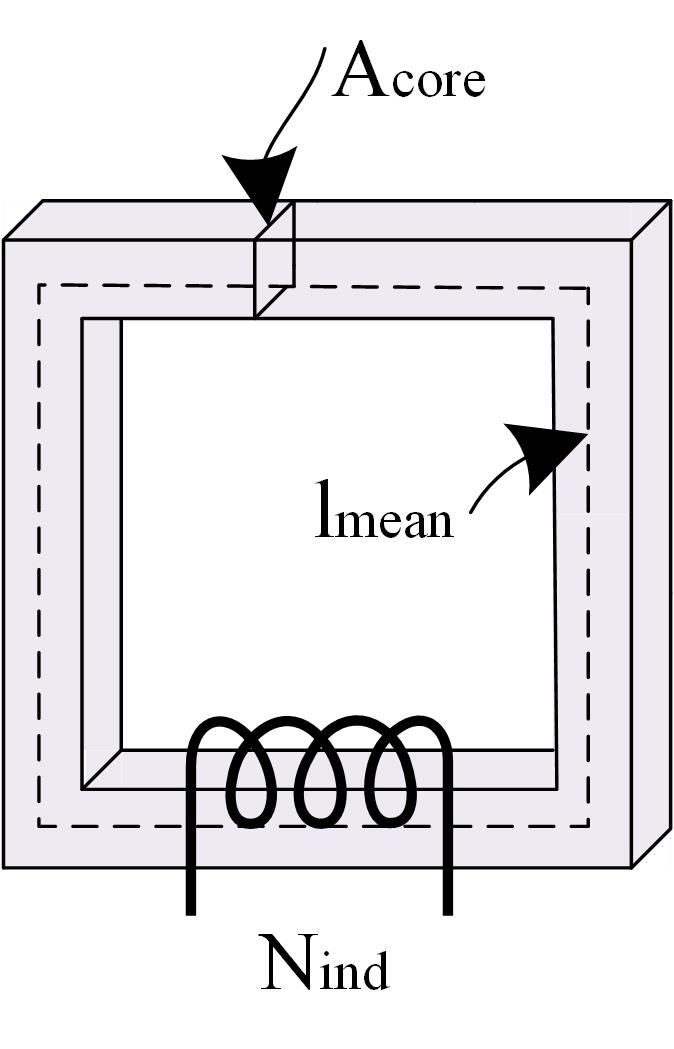}
\caption{Nonlinear inductor}
\label{fig:Fig. 2.2}
\end{figure}

\begin{figure}
\centering
\includegraphics[height=5cm]{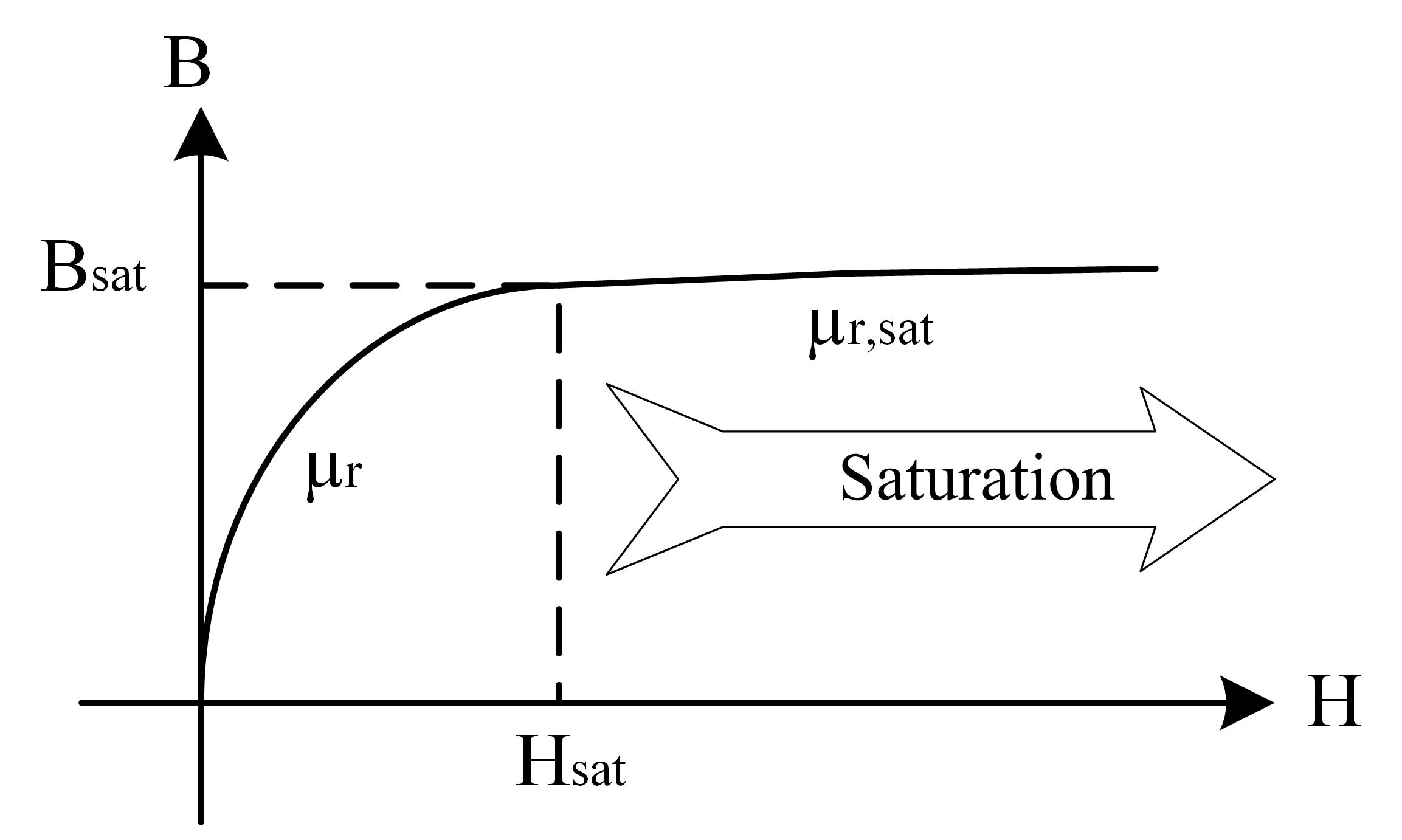}
\caption{Nonlinear B-H (hysteresis) curve in an iron core}
\label{fig:Fig. 2.3}
\end{figure}

The nonlinear inductor impedance is determined by the average longitude of the flux direction, $l_{mean}$, the cross-sectional area of the core, $A_{core}$, winding’s turn number, $N_{ind}$, and the core's relative permeability, ${\mu}_{r}$ (Fig.~\ref{fig:Fig. 2.2}). By conducting the core to a saturation mode, the inductance is smaller when the hysteresis curve’s operational point is outside of the saturation zone (Fig.~\ref{fig:Fig. 2.3}) \cite{5415715}.

The estimated inductances for saturated and unsaturated situations are given in the following equations:

\begin{equation}\label{5}
    {L_{x} = \mu_{0}\mu_{r} \times \frac{{N^2_{ind}} A_{core}}{l_{mean}}}
\end{equation}

\begin{equation}\label{6}
    {L_{y} = \mu_{0}\mu_{r,sat} \times \frac{{N^2_{ind}} A_{core}}{l_{mean}}}
\end{equation}

\begin{figure}
\centering
\includegraphics[height=9cm]{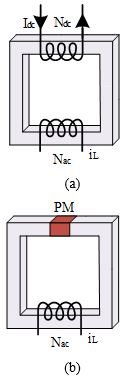}
\caption{Saturation methods of FCL core. (a) by direct current (b) by PM}
\label{fig:Fig. 2.4}
\end{figure}

\begin{figure}
\centering
\includegraphics[height=5cm]{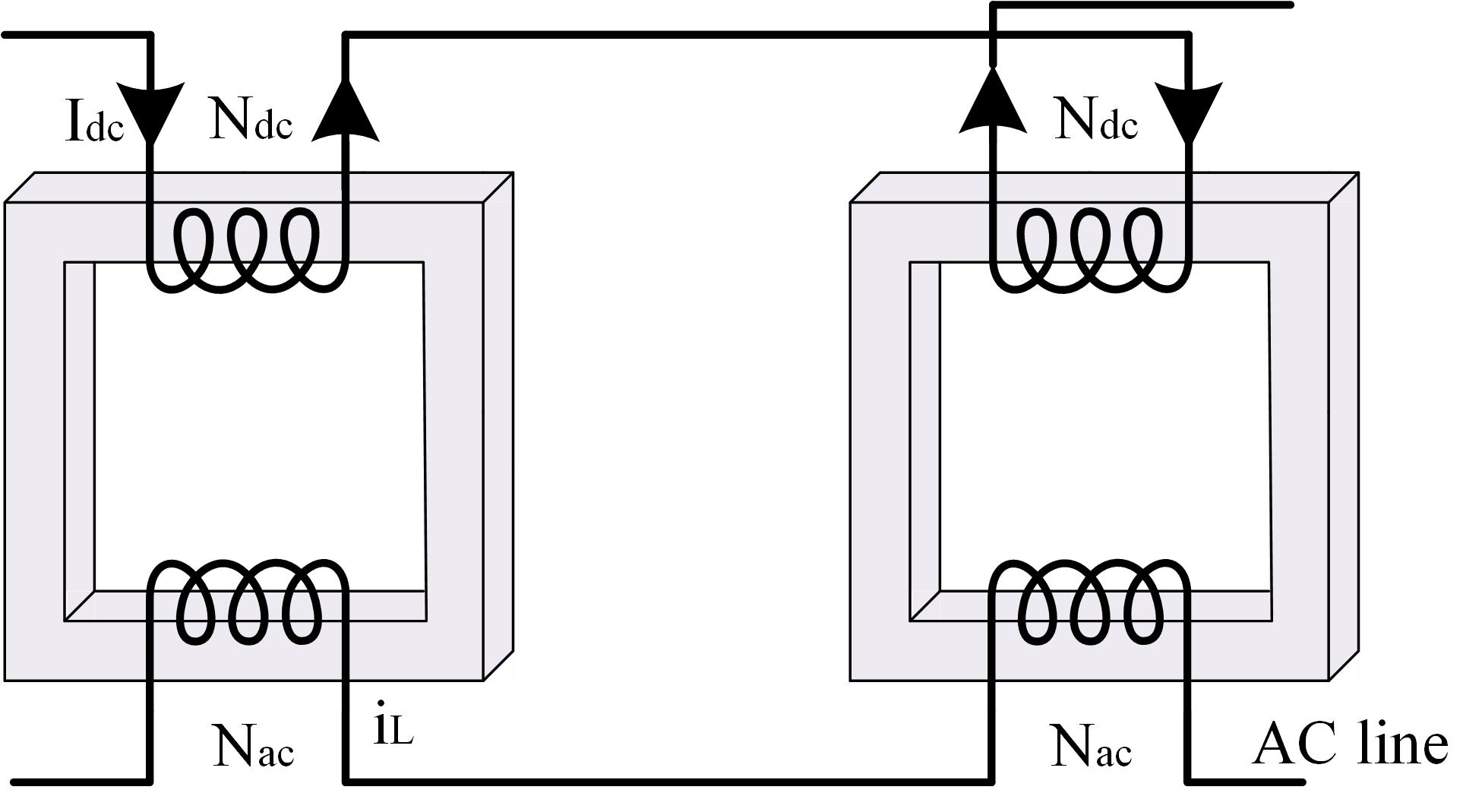}
\caption{Schematic of main components of SCFCLs}
\label{fig:Fig. 2.5}
\end{figure}

Where $L_{x}$ and $L_{y}$ are the FCL impedances during the fault and normal conditions, and $\mu_{0}$ and $\mu_{r,sat}$ are the air permeability and relative permeability coefficients of the core at a saturation condition. The core must be saturated under normal system conditions, but not over-saturated. The core should be de-saturated under fault situations. An additional dc current-carrying coil provided by an auxiliary source, or a permanent magnet (PM) can be used to saturate the core, as illustrated in Fig.~\ref{fig:Fig. 2.4}.\\
A dc coil, a copper ac coil, and an iron core make up generally the saturated core fault current limiters, the arrangement of which changes depending on the different constructions and used materials. The ac coils are linked to a power system in series, and the resulting magnetic field is directed in opposite directions. The dc coils are connected to a separate direct current source, as demonstrated in Fig.~\ref{fig:Fig. 2.5}.

The SCFCL works in its usual non-limiting state under normal power system circumstances, which is also the mode of operation for most of the time and has no impacts on a power grid. The rated current will flow between the ac winding windings, and the huge current in the dc winding will make the bias of the large dc magnetic field within a core in this case. During each ac cycle, deep saturation and low magnetic permeability are maintained in iron cores. The SCFCL reactance is low and has slight impact on the rest of the power system because of the inductance relation with the permeability \cite{jia2017numerical,cvoric2009new,gunawardana2016transient,nikulshin2016saturated}.

\begin{figure}
\centering
\includegraphics[height=8cm]{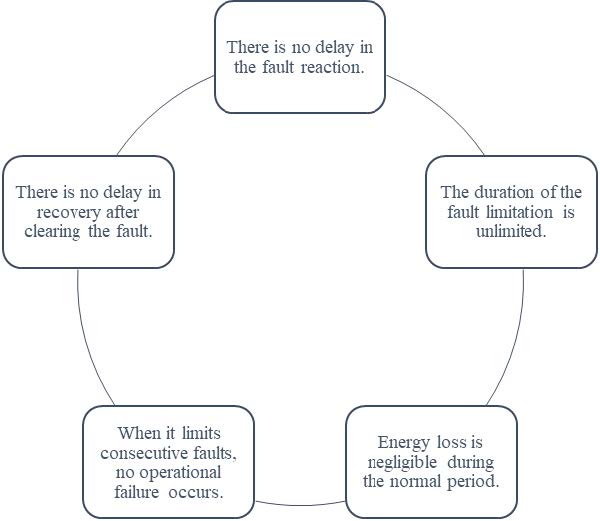}
\caption{SCFCL characteristics}
\label{fig:Fig. 2.6}
\end{figure}

A substantial fault current flows between the ac windings when short circuits occur, and the limiter enters its own current limiting mode. The dc current is instantly shut off for active type SCFCLs, and regardless of the short circuit fault, the same dc bias is created for inactive type SCFCLs. The iron cores in both types will not always be at deep saturation for a long period due to the high fault currents. The permeability rises by several thousand while the cores operate in the linear area of the hysteresis characteristic, therefore the average limiter reactance enhances. The limiter can restrict the fault current due to reactance \cite{chen2016parameter,baimel2021new}. Soft iron has nonlinear magnetic characteristics, which are dictated by the hysteresis characteristic of material and is used to represent in this case. The following graph (Fig.~\ref{fig:Fig. 2.6}) illustrates some advantages of employing this type of limiter.

\begin{figure}
\centering
\includegraphics[height=7cm]{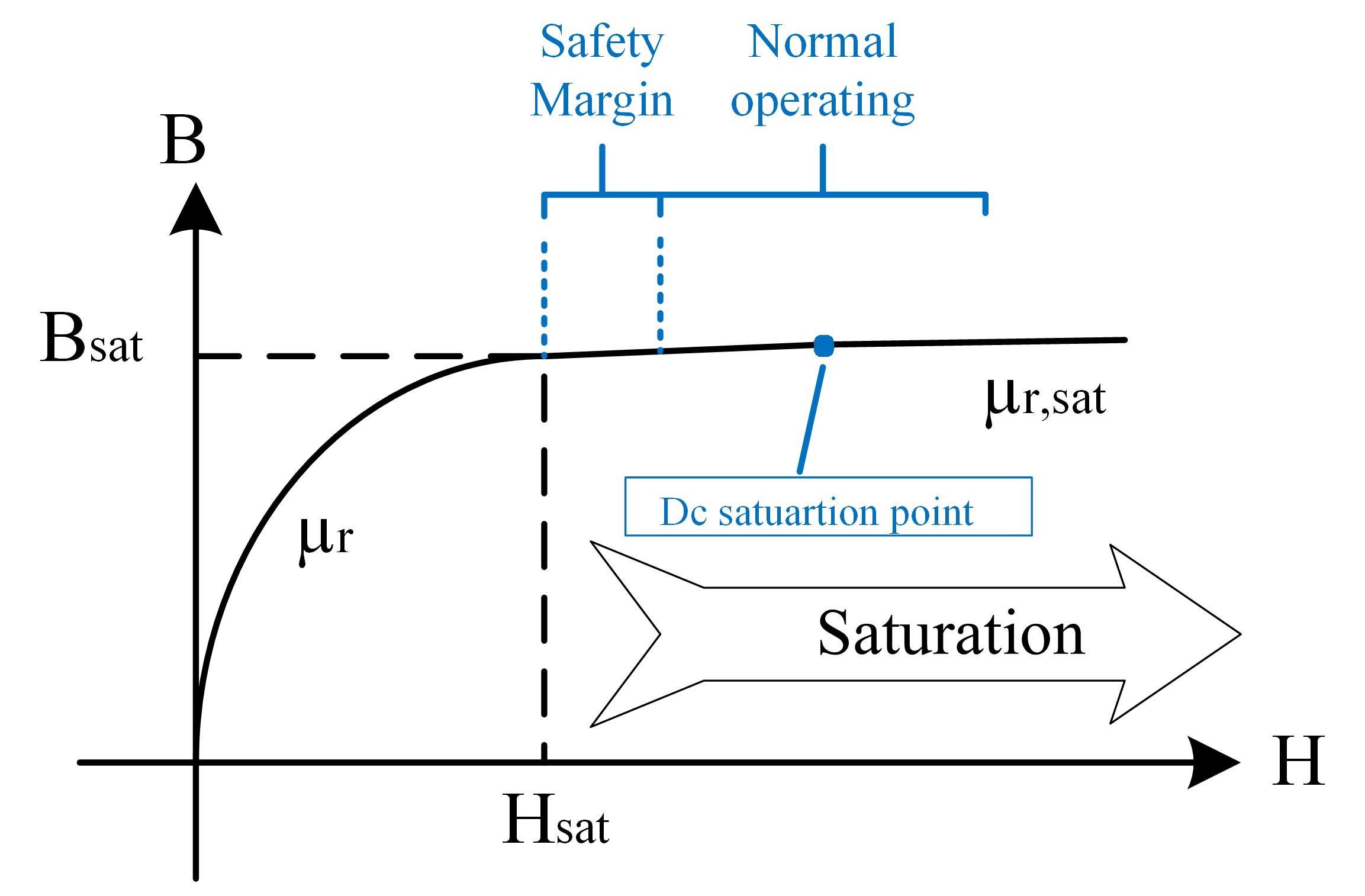}
\caption{FCL performance during normal regime – flux changes}
\label{fig:Fig. 2.7}
\end{figure}

The cores of limiters are adequately saturated to the point that typical current flux changes cannot restore them to the linear area depicted in Fig.~\ref{fig:Fig. 2.7}. The operating point of \textbf{B-H} graph is de-saturated in the case a fault occurrence, increasing the limiter impedance. The safe margin width indicated in the figure determines the fault flow level at which the FCL commences to restrict. The width of this portion may be ignored due to the small fault current peak, and the FCL can serve on spot.

The relationship between dc current, $I_{dc}$, line current, $I_{L}$, and the magnetic field in the core, $H$ is as follows. To calculate the safe zone width, $H_{s}$, Eq. (\ref{7}) must be rewritten as Eq. (\ref{8}) \cite{cvoric20163d}:

\begin{equation}\label{7}
    {N_{dc} I_{dc} - N_{ac} I_{L,max} > H_{sat} l_{mean}}
\end{equation}
\begin{equation}\label{8}
    {N_{dc} I_{dc} - N_{ac} I_{L,max} = (H_{sat} + H_{s}) l_{mean}}
\end{equation}

After the fault has been cleared, the core comes back to saturated mode. Therefore, when the amount of transmission current decreases, the ac winding’s impedance decreases with no latency. As a result, the inductive FCL can limit a countless number of continuous faults.\\
Because of the inductive essence of the fault current limiter’s impedance, there have been no energy losses throughout the restriction, but there is a loss owing to the resistance of a winding. When compared to the transmission line's nominal power, the dissipative power is rather low. Unlike resistive fault current limiters, the limiting operation period is not constrained by an acceptable quantity of thermal transfer at impedance. Because inductive limiters can't operate as a breaker, the fault current must be eliminated by other protection apparatus like circuit breakers. However, attributed to the design issues, they have not yet been industrialized \cite{yuan2020saturated}: 1- The scale and expense of the materials are both significant, 2- Inductive overvoltage passing through the dc current source during fault operation.

\section{Circuit Diagram of a Simple Power System}
\begin{figure}
\centering
\includegraphics[height=2cm]{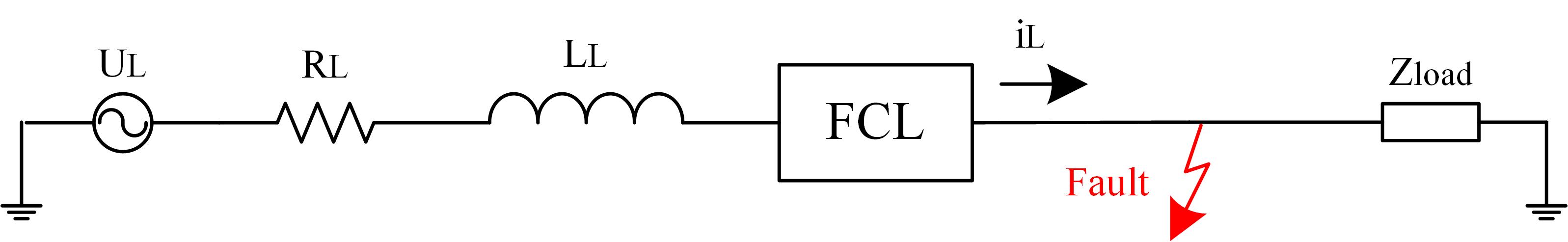}
\caption{Power system equivalent circuit with single phase to ground fault}
\label{fig:Fig. 2.8}
\end{figure}

Test circumstances of FCLs and their impacts on current limiting and other network parameters are in the form of a simplified power network test circuit with resistance, inductor, power supply, and load resistance, which act in fault and normal operations utilizing the various architectures of SCFCLs. The sample circuit system associated with their values are illustrated in Fig.~\ref{fig:Fig. 2.8} and Table \ref{table 2.1}. It is our purpose to test the different type of saturated core FCLs on the sample network by providing a model by finite element method simulated by COMSOL Multiphysics software in a computer system with Intel core i7, 6500U, 64-bit, 12 GB RAM and provide a comparison between them in terms of appropriate parameters. Various arrangements of these FCLs are given in the next section.

\section{Different Configurations of SCFCLs}
We propose some SCFCL configurations such as 100\% magnetic separation, partial magnetic separation, short circuit dc winding, etc. The performance of these FCLs is tested on the sample power system, and comparisons are provided.

\begin{table}
\centering
%\tiny
\caption{Equivalent circuit parameters}

\label{table 2.1}
%\begin{center}
\setlength{\tabcolsep}{10pt} % Default value: 6pt
\renewcommand{\arraystretch}{1.5} % Default value: 1
\begin{tabular}{|c|c|}
\hline
 \textbf{Parameter} & \textbf{Description} \\ \hline
Power Supply ($U_L$) (rms) & $10\sqrt{2} kV$  \\ \hline
$R_L$ & $0.1095 \Omega$  \\ \hline
$L_L$ & $5.63419 \times 10^{-4} H$  \\ \hline
$Z_{load}$ & $8.79 \Omega$  \\ \hline
\end{tabular}
%\end{center}
\end{table}

\subsection{SCFCL with 100\% Magnetic Separation}
Because of resistive essence of ac winding in SCFCLs, there will be a voltage drop which is a coefficient of voltage during normal operation (up to 2\%). However, phase voltage of $U_{L(t)}$, is applied to the ac winding while a fault occurs. An overvoltage $(U_{FCL,dc(t)})$ is produced by a dc source as a result of windings’ coupling in a transformer in which the number of turns in the ac and dc windings are $N_{ac}$ and $N_{dc}$, correspondingly \cite{li2019current,wei2018performance}.

\begin{equation}\label{9}
    {U_{FCL,dc (t)} = \frac{N_{dc}}{N_{ac}} \times U_{L} (t)}
\end{equation}

\begin{figure}
\centering
\includegraphics[height=3.5cm]{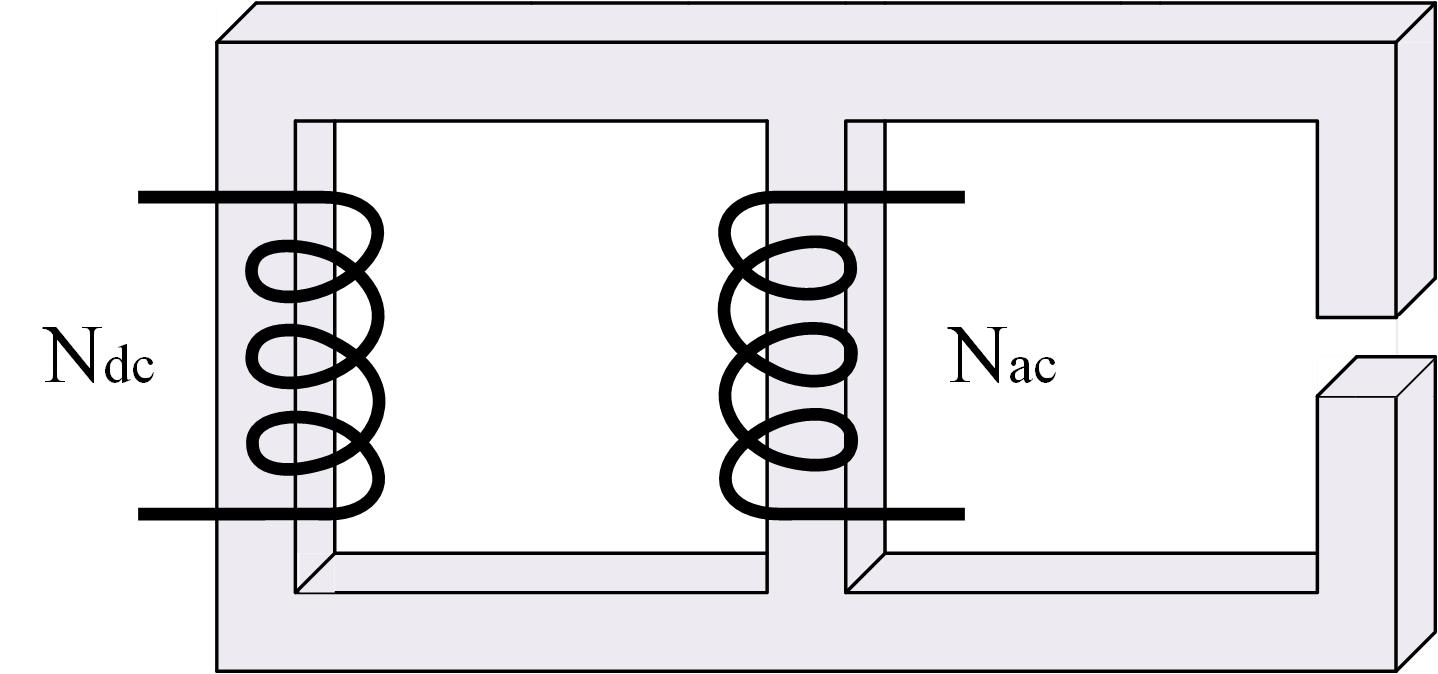}
\caption{SCFCL configuration with 100\% magnetic separation of dc and ac circuit}
\label{fig:Fig. 2.9}
\end{figure}

The induced overvoltage problem can be solved by magnetic separation of ac and dc windings. Fig.~\ref{fig:Fig. 2.9} depicts this arrangement in this case.

In this situation, the dc flux only crosses over the core's left and center legs and reaches saturation there. A leg with a significant reluctance possessing an air gap deflects the passing flux away from the path as Eq. (\ref{10}):

\begin{equation}\label{10}
 \left.\begin{aligned}
        \Re_{m} = \frac{l_{mm}}{\mu_{0}\mu_{r}A_{cm}} \\
        \Re_{rr} = \frac{l_{mr} + \mu_{r} l_{g}}{\mu_{0}\mu_{r}A_{cr}}
       \end{aligned}
 \right\} \longrightarrow \Re_{m} << \Re_{rr} \quad \text{that} \quad 
  %\qquad \text{R_{m}  R_{rr}}
\end{equation}
\\
\\
{
\centering
- $\Re_{m}$ is middle leg reluctance \\
- $\Re_{rr}$ is right leg reluctance \\
- $l_{mm}$ is average length of middle leg \\
- $l_{mr}$ is average length of right leg \\
- $l_{g}$ is air gap length \\
- $A_{cm}$ is cross-section of middle leg \\
- $A_{cr}$ is cross-section of right leg \\}

\vspace{5mm}
During the fault condition, the ac flux merely flows through the center and right legs which is induced by fault currents. The reluctance of left side is much higher than right one (with air gap) owing to a saturation state \cite{cvoric20163d}:

\begin{equation}\label{11}
 \left.\begin{aligned}
        \Re_{lf} = \frac{l_{mf}}{\mu_{0}\mu_{r}A_{cl}} \\
        \Re_{rr} = \frac{l_{mr} + \mu_{r} l_{g}}{\mu_{0}\mu_{r}A_{cr}}
       \end{aligned}
 \right\} \longrightarrow \Re_{lf} >> \Re_{rr} \quad \text{that} \quad 
  %\qquad \text{R_{m}  R_{rr}}
\end{equation}
\\
\\
{
\centering
- $\Re_{lf}$ is left leg reluctance \\
- $l_{mf}$ is average length of left leg \\
- $A_{cl}$ is cross-section of left leg \\}

\vspace{5mm}

\begin{figure}
\centering
\includegraphics[height=4cm]{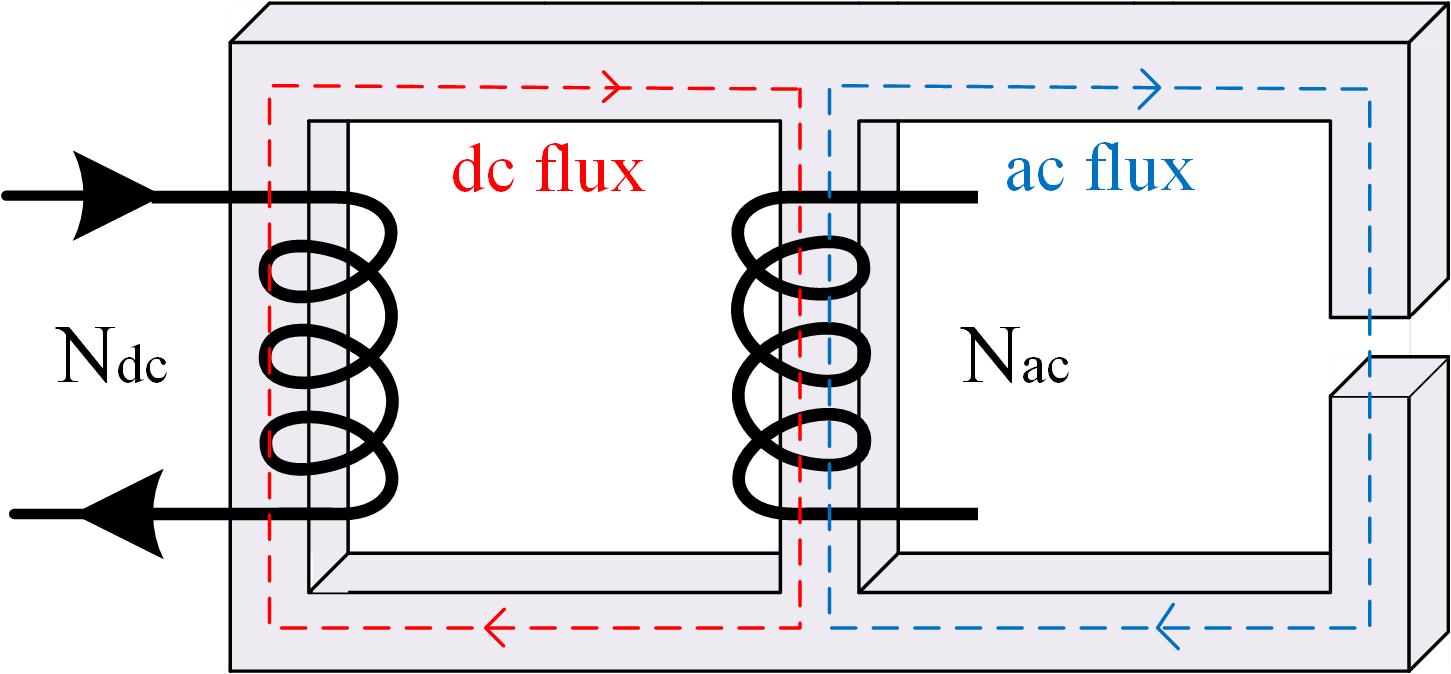}
\caption{Flux distribution in SCFCL with 100\% magnetic separation}
\label{fig:Fig. 2.10}
\end{figure}

The dc winding does not sense any flux change in this situation since it is not aware of the ac flux. As a result, no induced overvoltage occurs. Fig.~\ref{fig:Fig. 2.10} depicts the flow distribution, with two cores utilized for each phase.

This configuration, however, has a design flaw. During normal operation, the ac winding has a very high impedance. Non-deep saturation of the mid-leg results in high impedance. Because the dc winding is on the opposite leg of the core, it can only saturate the same leg.

\subsection{SCFCL with Partial Magnetic Separation}
\begin{figure}
\centering
\includegraphics[height=5cm]{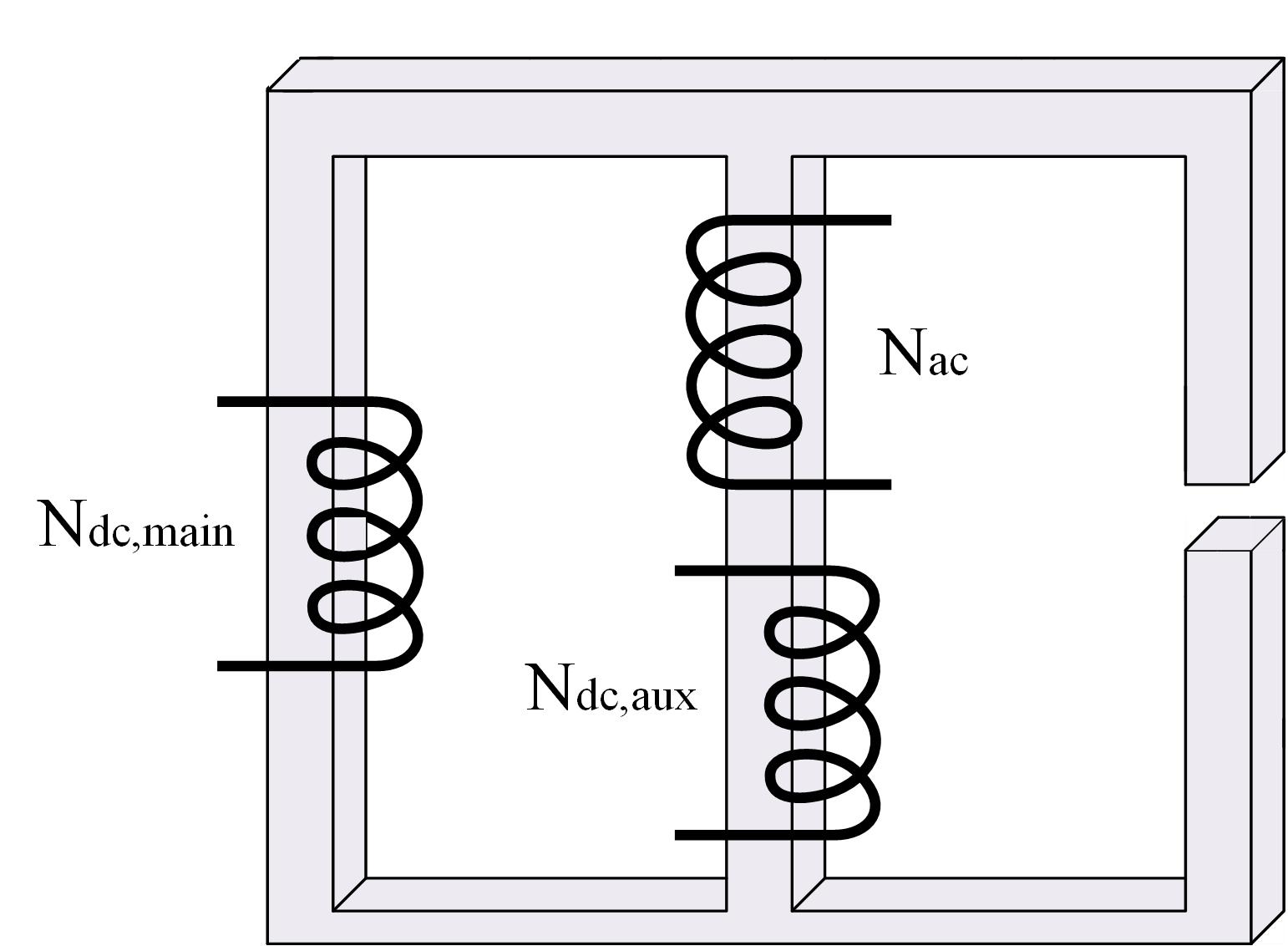}
\caption{SCFCL with partial magnetic separation}
\label{fig:Fig. 2.11}
\end{figure}

\begin{figure}
\centering
\includegraphics[height=5.5cm]{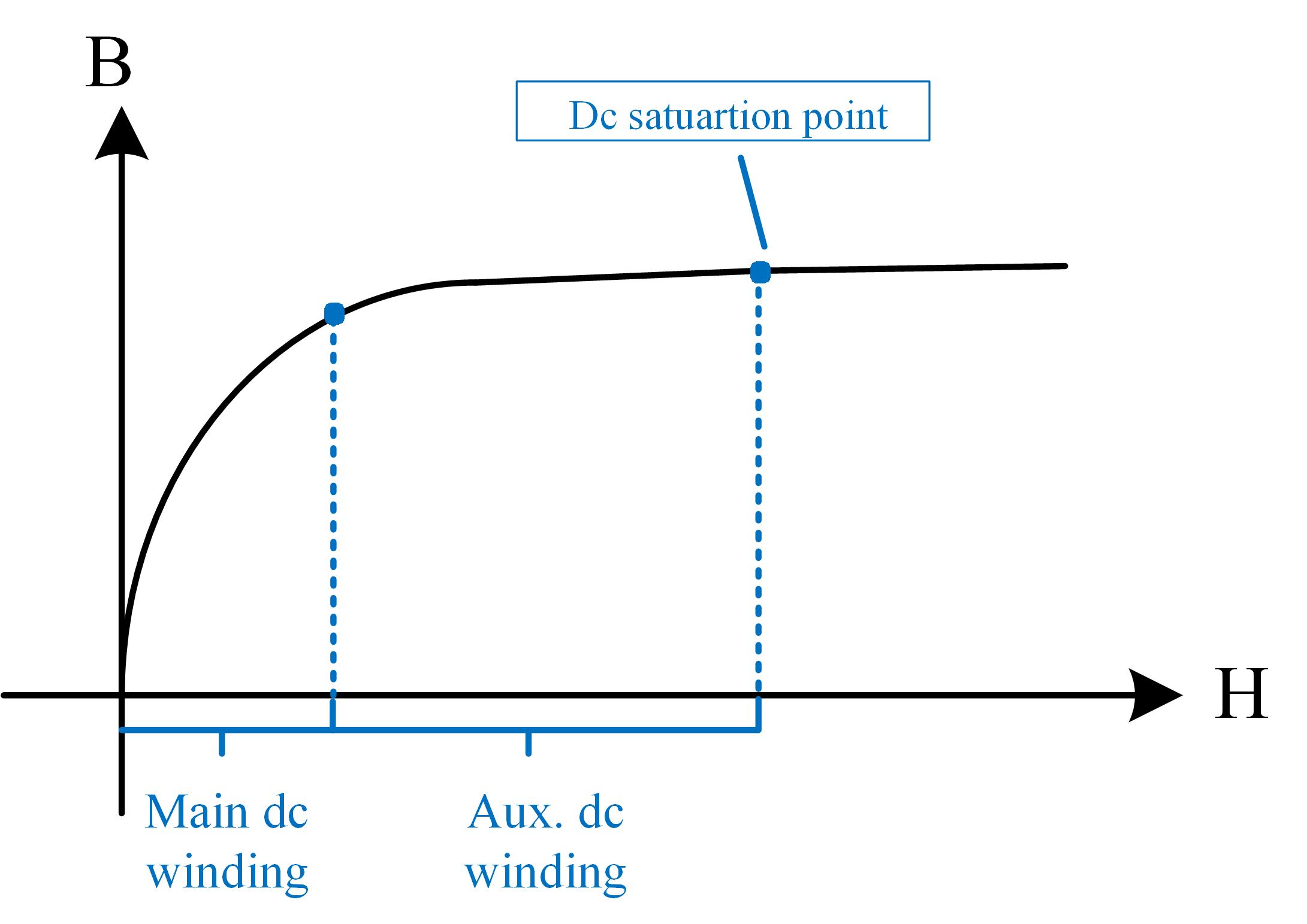}
\caption{Deep saturated mode of an AC leg provided by an dc auxiliary coil}
\label{fig:Fig. 2.12}
\end{figure}

The saturated core FCL with 100\% magnetic separation has a limiting difficulty in that a coil placed in the opposite leg of the core cannot be thoroughly saturated, as detailed in the preceding section. The ac leg falls out of saturation even with a regular current, increasing the FCL impedance during normal operation. The core size must be large enough to get an acceptable low value for the FCL's normal state impedance, it may be inferred \cite{naphade2021experimental}.\\
An improved design of a SCFCL with partial magnetic separation is illustrated in Fig.~\ref{fig:Fig. 2.11}. There is an ac leg with an appended dc auxiliary coil to conduct it to a stronger saturated mode. The effect of the auxiliary dc coil is depicted in Fig.~\ref{fig:Fig. 2.12}. This auxiliary coil propels the working point outside the knee area, allowing it to saturate better. SCFCL parameters are presented in Table \ref{table 2.2}.

\begin{table}
\centering
%\tiny
\caption{SCFCL parameters}

\label{table 2.2}
%\begin{center}
\setlength{\tabcolsep}{3.5pt} % Default value: 6pt
\renewcommand{\arraystretch}{1.1} % Default value: 1
\begin{tabular}{|c|c|c|c|}
\hline
 \textbf{Parameter} & \textbf{Description} & \textbf{Value} & \textbf{Value}\\ \hline
& & 100\% magnetic separation & Partial magnetic separation  \\ \hline
$N_{ac}$ & No. of turns in ac winding & 60 & 60  \\ \hline
$N_{dc}$ & No. of turns in dc winding & 500 & 500  \\ \hline
$I_{dc}$ & dc current & 450 & 450  \\ \hline
$l_{gap}$ & air gap length & 0.3 & 0.3  \\ \hline
$N_{dc,aux}$ & No. of turns in auxiliary winding & - & 76  \\ \hline
\end{tabular}
%\end{center}
\end{table}

Eq. (\ref{12}) illustrates a relationship between $N_{ac, aux}$ and $N_{ac}$ as follows:

\begin{equation}\label{12}
 N_{dc,aux} I_{dc} = N_{ac} I_{L} \quad \text{that} \quad I_{dc} = \text{dc current}, \quad I_{L} = \text{Line current}
\end{equation}

The design parameters are analogous to the previous mode for this case and the auxiliary winding has 76 turns, which is obtained using Eq. (\ref{12}). The FCL also comes with the same specifications as the auxiliary dc coil, except that the main dc coil is short-circuited.

\subsection{SCFCL with DC Coil through short-circuited terminals}
\begin{figure}
\centering
\includegraphics[height=5cm]{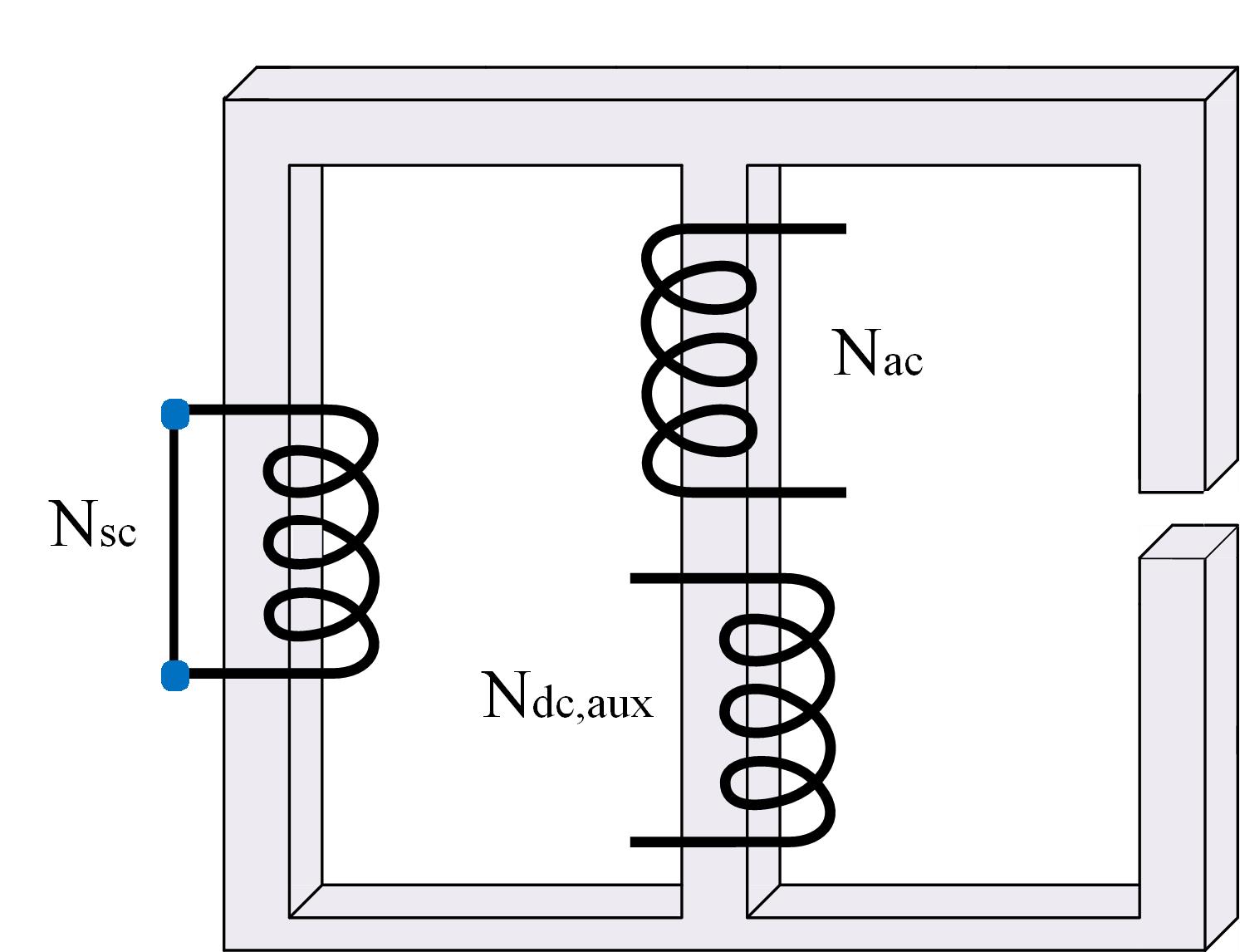}
\caption{SCFCL with DC Coil through short-circuited terminals}
\label{fig:Fig. 2.13}
\end{figure}

Even though the primary dc coil has more turns than the auxiliary one, the primary coil’s share of ac leg saturation is less than that of the auxiliary dc coil illustrated in Fig.~\ref{fig:Fig. 2.13}. To raise the dc reluctance and control the ac flux via the air gap, more turns of the dc primary coil are indispensable. Otherwise, the induced voltage at the dc source will rise due to the spreading of ac flux via the dc leg \cite{wei2020limiting,yuan2020saturated}.\\
If the primary winding is cut off from the rest of the circuit, the limiter functionality will not change as shown in the figure below. The ac current flowing via the dc leg causes a current in the coil with short-circuited terminals, which drives the ac flux. The dc leg is not fully saturated in this scenario. A supplementary coil navigates it to the knee area. However, the FCL impedance does not increase when everything is running smoothly because it is where the ac flux flows.

The level of flux variation in the dc side determines the induction current in the short-circuit coil, which is oppositely proportional to $N_{dc}$. The quantity of induction current increases as a reduction of the coil's turn numbers. Considering that, the entire amount of material could be substantially reduced in this circumstance. The overall amount of FCL coil material is lowered by substituting the primary dc coil with a short-circuit one by partial magnetic separation \cite{cvoric2009new}.

\section{Simulation Results for Configurations with One AC Winding}
In this section, we evaluate the simulation results of these models, which we define the models as Fig.~\ref{fig:Fig. 2.14} and the components of this type of limiter as Fig.~\ref{fig:Fig. 2.15}:

\begin{figure}
\centering
\includegraphics[height=4cm]{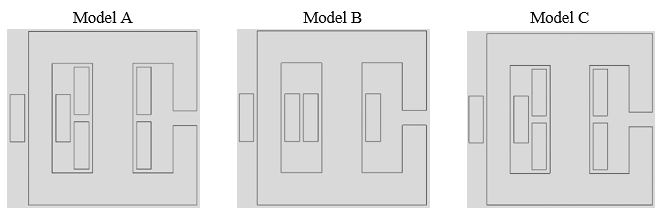}
\caption{SCFCL structures including partial magnetic separation (Model A), 100\% magnetic separation (Model B) and a dc coil with short-circuited terminals (Model C)}
\label{fig:Fig. 2.14}
\end{figure}

\begin{figure}
\centering
\includegraphics[height=5.5cm]{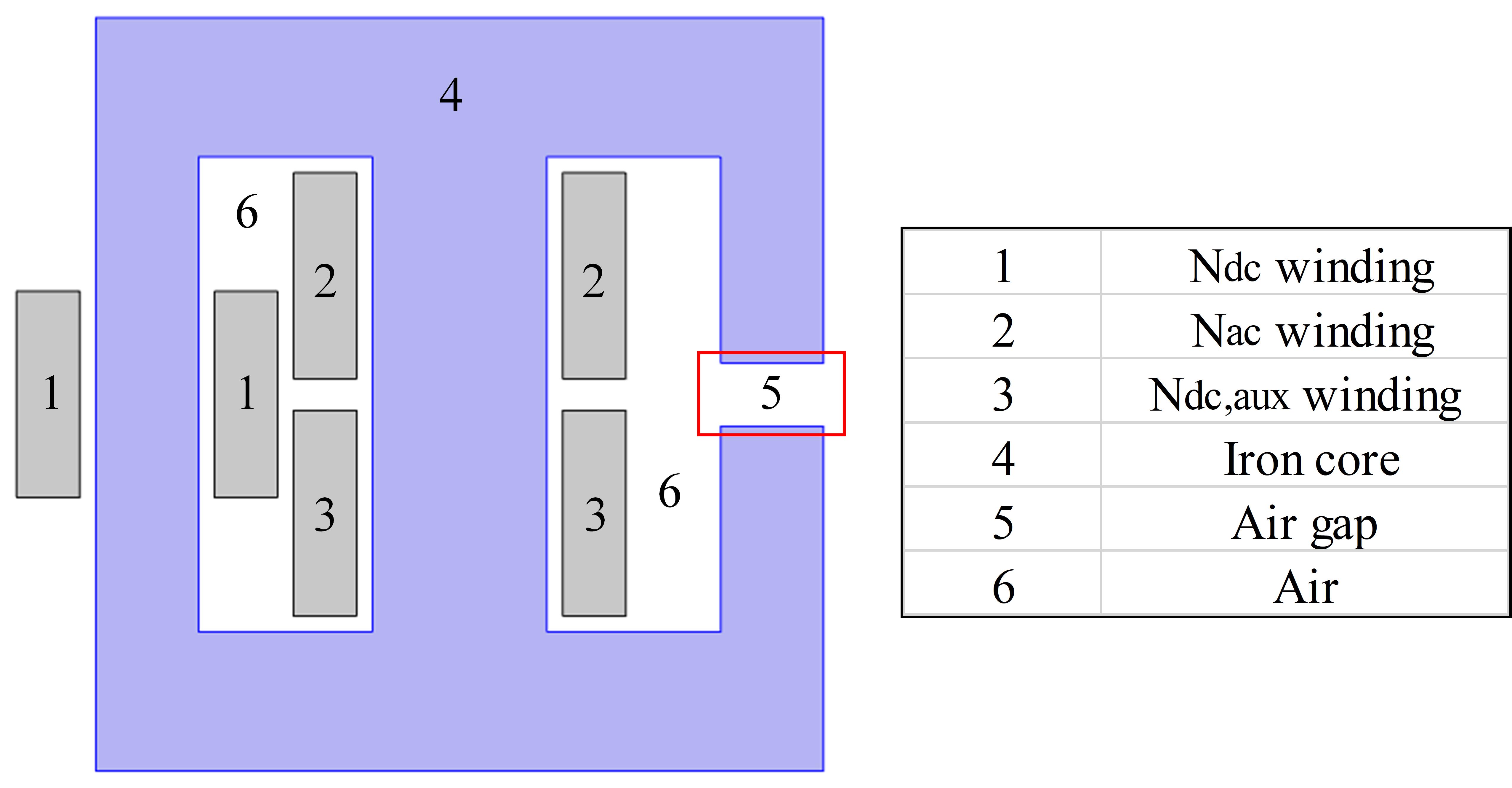}
\caption{SCFCL components}
\label{fig:Fig. 2.15}
\end{figure}

Also, the geometrical dimensions of the design are demonstrated in Fig.~\ref{fig:Fig. 2.16}:

\begin{figure}
\centering
\includegraphics[height=6cm]{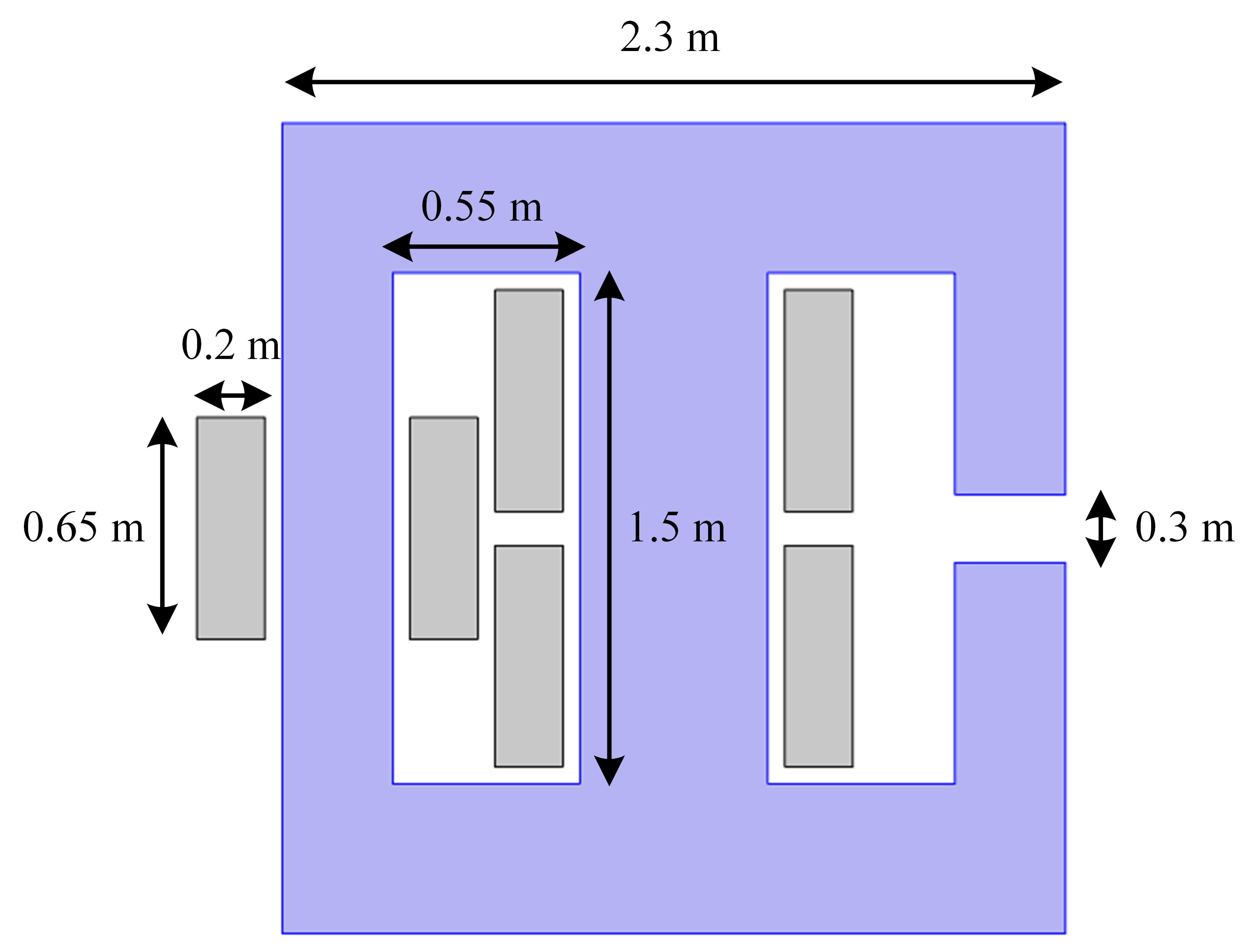}
\caption{Configuration dimensions}
\label{fig:Fig. 2.16}
\end{figure}

\begin{figure}
\centering
\includegraphics[height=6cm]{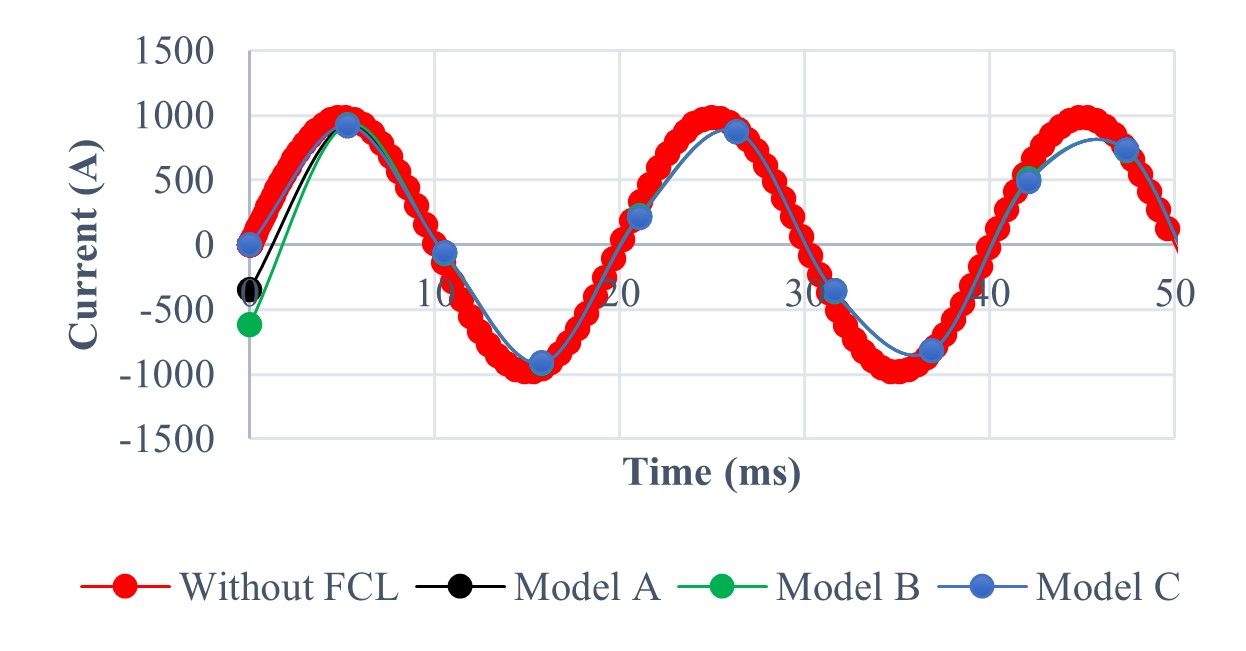}
\caption{Current passing through the system in normal operating conditions}
\label{fig:Fig. 2.17}
\end{figure}

Using the sample power system for assessing the FCL performance in the circuit in fault conditions, the current curves passing through the system and the flux density and field strength curves can be obtained, as well as the distribution of flux density or field strength in different situations. A transient analysis was performed for the configurations versus time and its performance in the presence of fault is compared and assessed. Even though the limiter exists in a power system during normal situation, it should not have impacts on the circuit. Therefore, the current passing through the grid without the presence of FCL and with the presence of FCL under normal operating conditions of the system is compared, which is shown in Fig.~\ref{fig:Fig. 2.17}. Red line demonstrated the electric current without FCL with amplitude of 1000 A. Other models illustrate some changes in current specifically at the outset of time. It originates from copper structure of FCL and is due to winding positions that reduces the amplitude of current at the models.

\begin{figure}
\centering
\includegraphics[height=6cm]{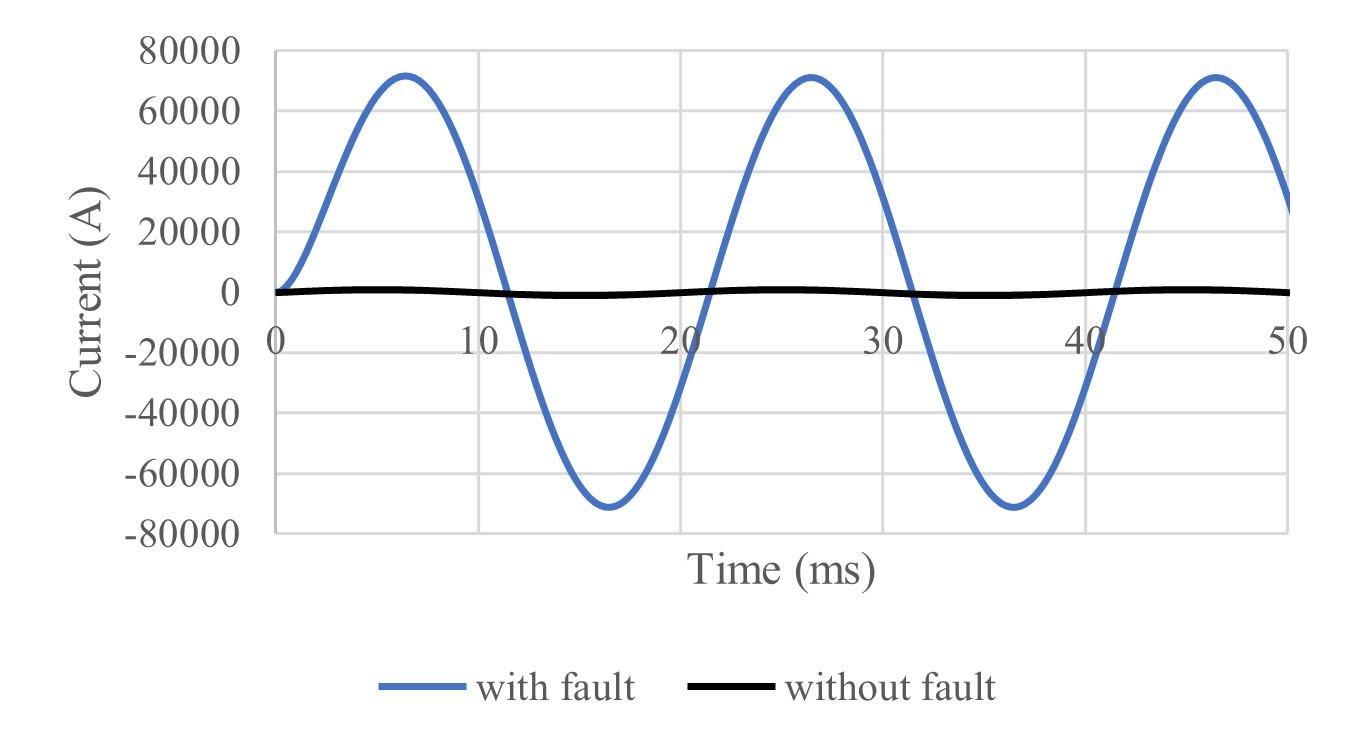}
\caption{Current passing through the system in normal and fault operation without SCFCL}
\label{fig:Fig. 2.18}
\end{figure}

\begin{figure}
\centering
\includegraphics[height=6cm]{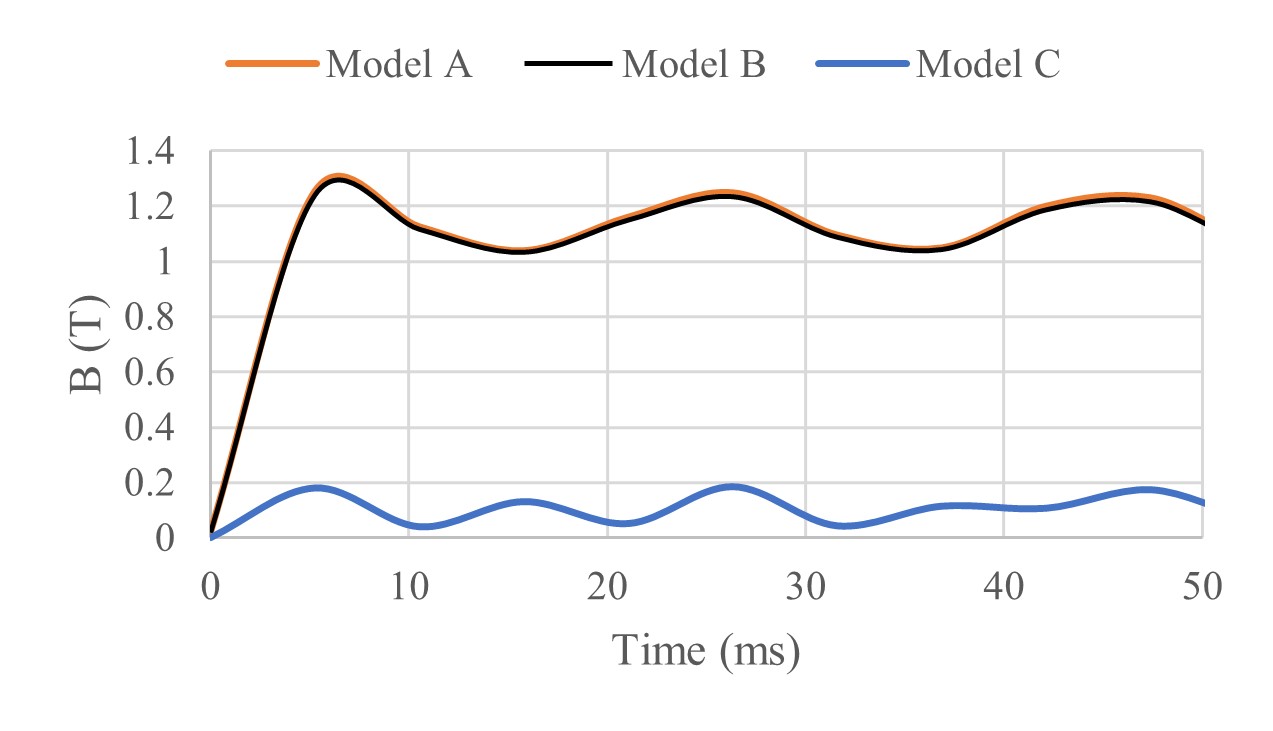}
\caption{Magnetic flux density changes versus time for different models}
\label{fig:Fig. 2.19}
\end{figure}

In models A, B and C, the limiter is in the electric circuit. Moreover, a maximum current passing through the system is about 1000 A, which with the presence of FCL (because of series connection to the circuit) due to a slight voltage drop (very small impedance input), the amount of current is slightly reduced. Also, the general diagram of the current without the presence of FCL under normal and fault operating conditions is illustrated in Fig.~\ref{fig:Fig. 2.18}, where the fault was initially entered into the system and the analysis was performed on the circuit for 50 ms. As you can observe, the fault impact on the electric current is highly significant.\\
Also, as mentioned, only the dc leg goes to deep saturation in model B, and to further saturate the ac leg, a dc auxiliary winding is added to it that a flux density diagram is shown in Fig.~\ref{fig:Fig. 2.19}. As you can observe, the flux density in Model A (partial mode) is slightly better saturated due to the auxiliary winding. Also in the Model C, the dc leg is not deeply saturated and is conducted by an auxiliary winding to the knee area, which has low values of flux density, as depicted in Fig.~\ref{fig:Fig. 2.19}.

\begin{figure}
\centering
\includegraphics[height=6.5cm]{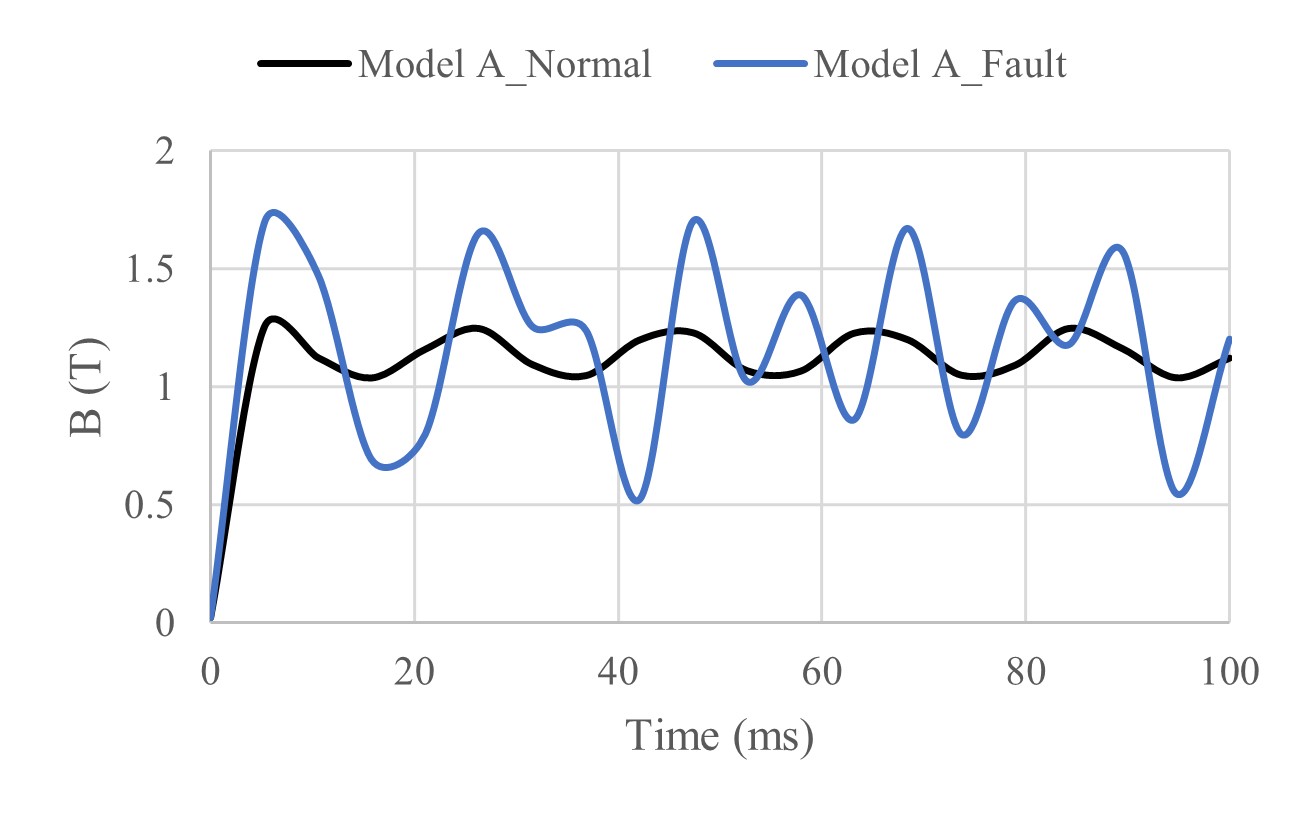}
\caption{Magnetic flux density passing through the core for both conditions in model A}
\label{fig:Fig. 2.20}
\end{figure}

Now, if we examine the flux density for model A in normal operation and fault mode, we perceive that the flux density in fault mode is generally higher than normal condition, which the fault causes more saturation of the core and a larger impedance. The flux density is shown in Fig.~\ref{fig:Fig. 2.20} that represents this type of SCFCL is effective in density reduction and the maximum amount of $B$ reaches almost 1.3 T.

\begin{figure}
\centering
\includegraphics[height=6.5cm]{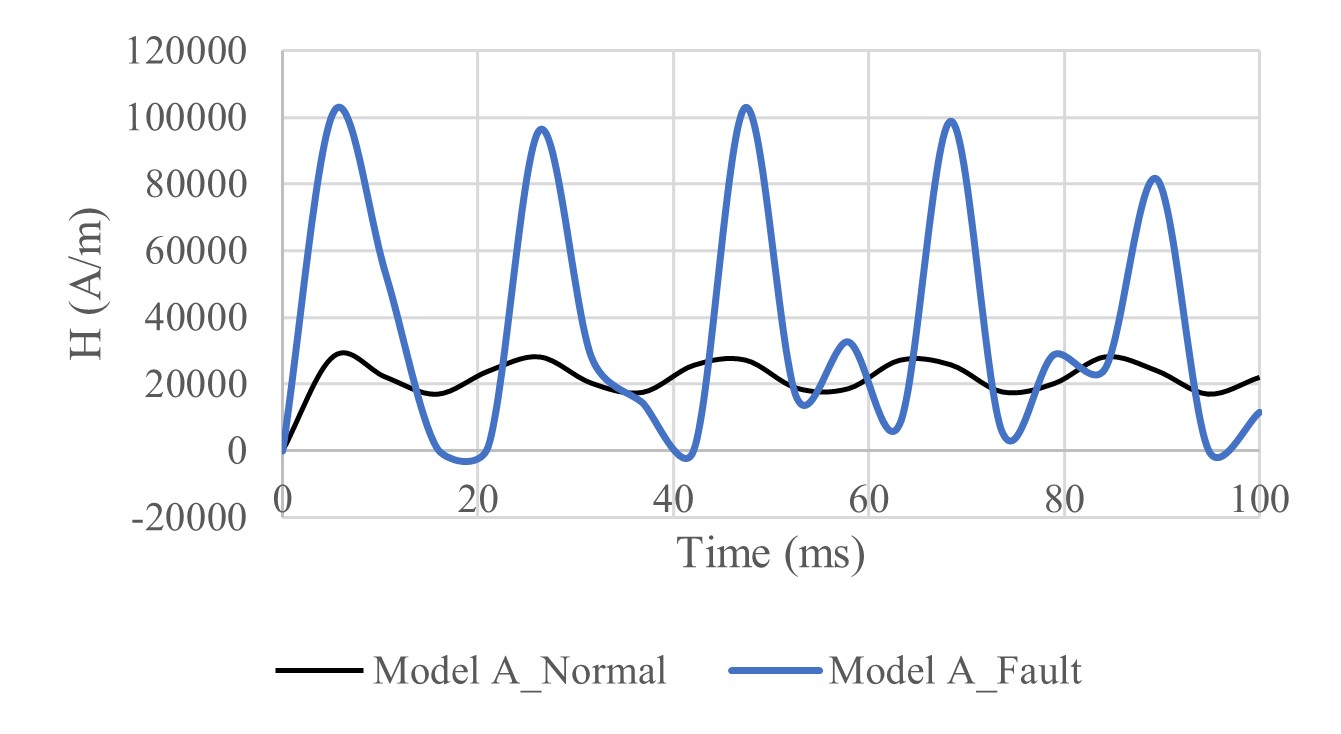}
\caption{Magnetic field strength passing through the iron core for model A in normal and fault modes}
\label{fig:Fig. 2.21}
\end{figure}

Fig.~\ref{fig:Fig. 2.21} depicts the magnetic field strength of the iron core for Model A as well, which is relatively higher in the fault state.

\begin{figure}
\centering
\includegraphics[height=6.5cm]{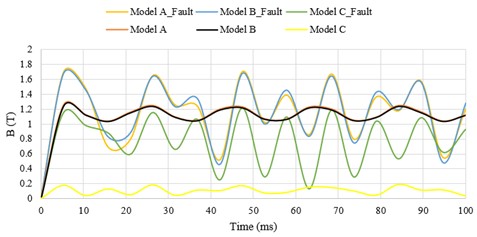}
\caption{Comparison of flux density in normal and fault operations}
\label{fig:Fig. 2.22}
\end{figure}

For a general comparison of flux density in the models presented above, their diagrams in normal operation and fault modes are represented in Fig.~\ref{fig:Fig. 2.22}. The lowest flux density value is related to dc short-circuited winding mode in normal operation which is in saturation mode. Most of the mode is related to the partial model in the fault mode, which possesses deeper saturation. Significant difference in normal and fault modes for Model C is due to the presence of short-circuited dc winding that the impact of this type of SCFCL is very advantageous. Other structures have smaller decline in flux density than Model C.

The amount of current flow in this model in fault mode has also been investigated in the presence of SCFCL. In this case we have\\

 $B = \mu H \xrightarrow[\text{}] {if} \quad B \uparrow \rightarrow H \uparrow, \qquad H = \frac{NI}{l} \xrightarrow[\text{}] {N} I (\text{variable}),$
\\
\\
Which can be analyzed using the flux density diagram.\\

\begin{figure}
\centering
\includegraphics[height=6.5cm]{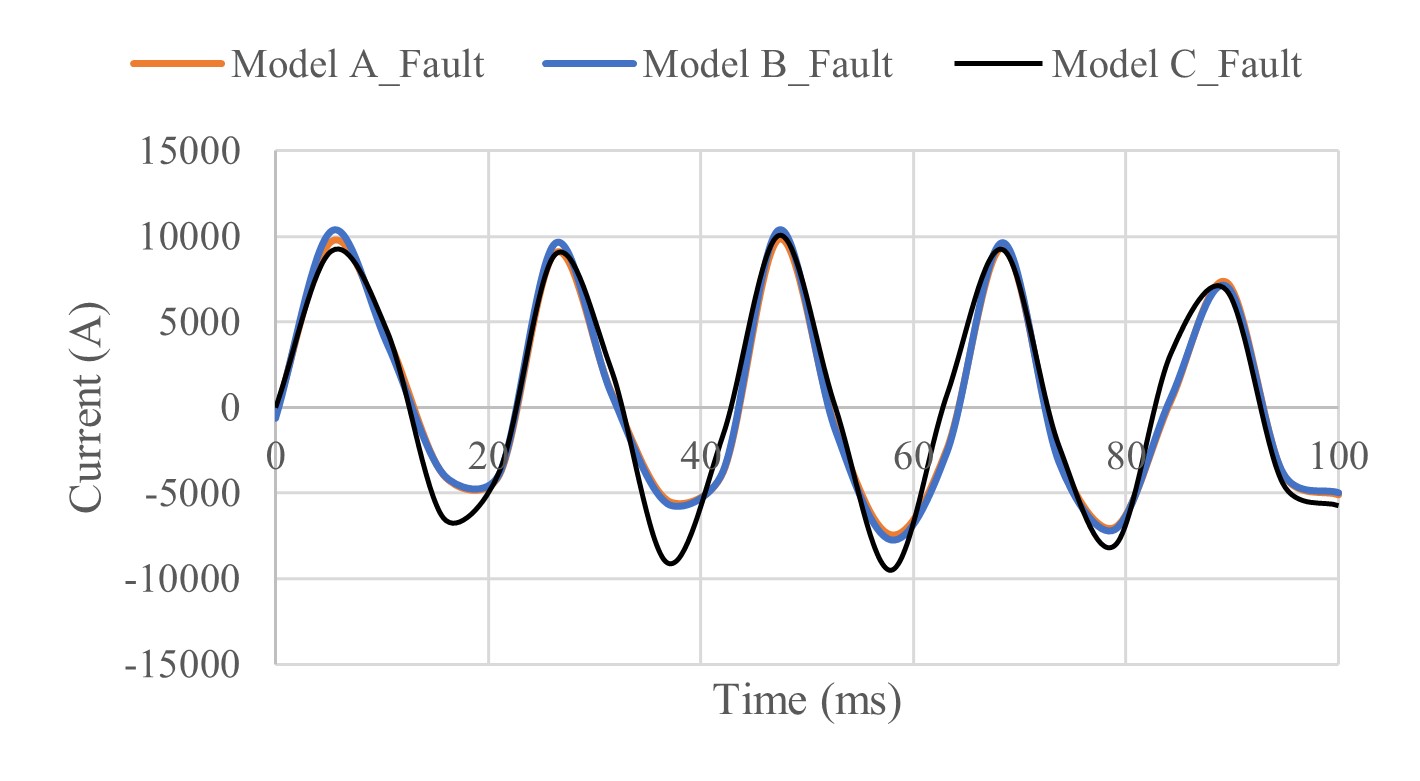}
\caption{Current passing through the system during a fault condition with SCFCL}
\label{fig:Fig. 2.23}
\end{figure}

\begin{figure}
\centering
\includegraphics[height=6.5cm]{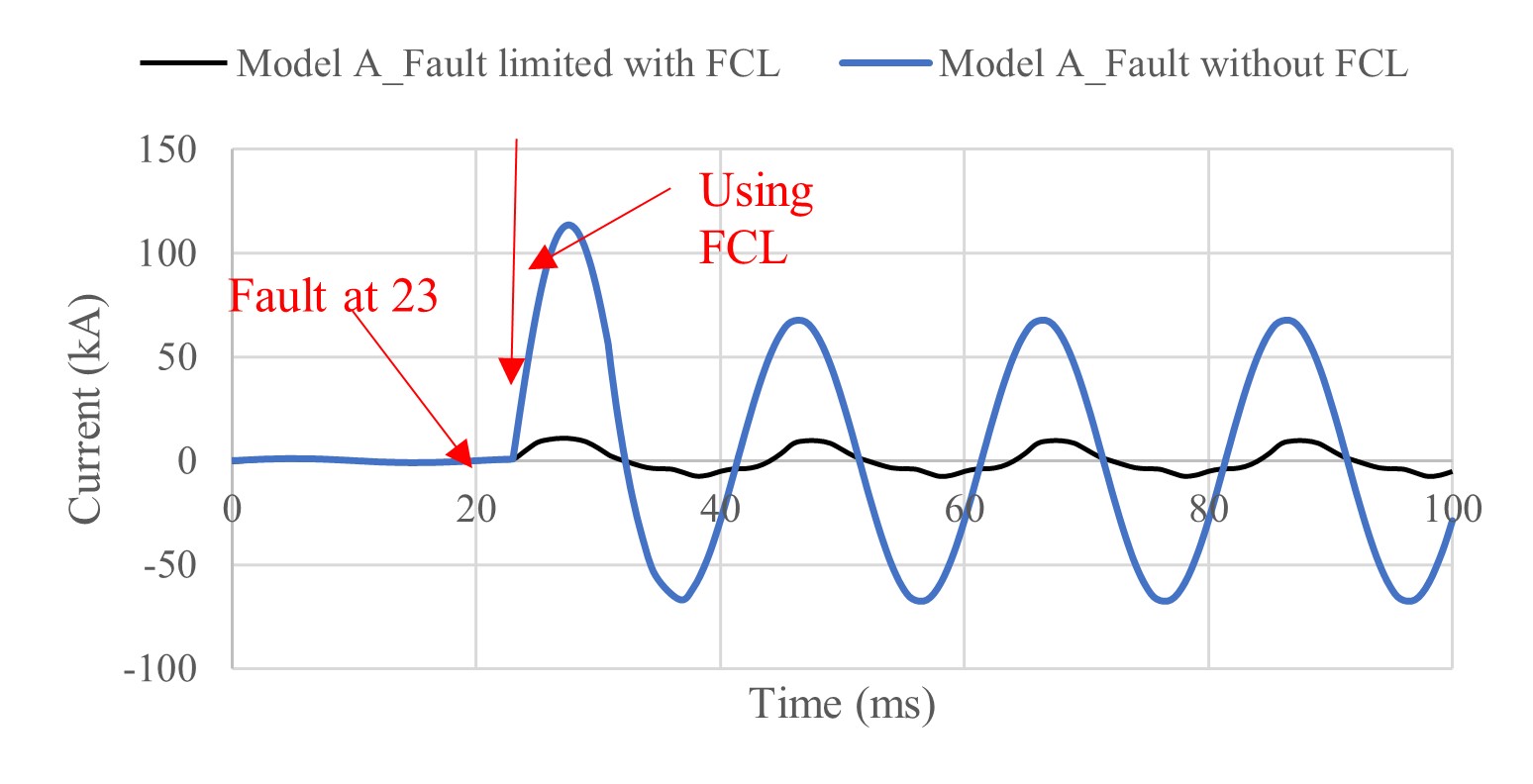}
\caption{Current passing through the system by applying a fault in 23 ms}
\label{fig:Fig. 2.24}
\end{figure}

For example, a fault has been occurred on 23 ms, and the FCL performance for Model A has been investigated, as shown in Fig.~\ref{fig:Fig. 2.24}. As you can see, there is no fault until 23 ms and the FCL is in the circuit, so it should not normally have much of an effect on the circuit, which makes sense. After starting the fault in the first peak, the current was almost 116 kA, and then the peak of fault current was reduced, which the FCL has amplified properly (we applied this type of fault on 23 ms in this section Because the calculation volume is very high and time consuming and for analysis and comparison, so a fault has been applied in this moment).

\begin{figure}
\centering
\includegraphics[height=8cm]{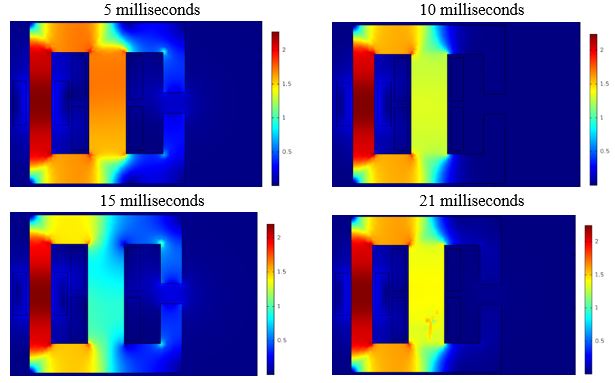}
\caption{Flux density illustration during normal mode for Model A}
\label{fig:Fig. 2.25}
\end{figure}

\begin{figure}
\centering
\includegraphics[height=8cm]{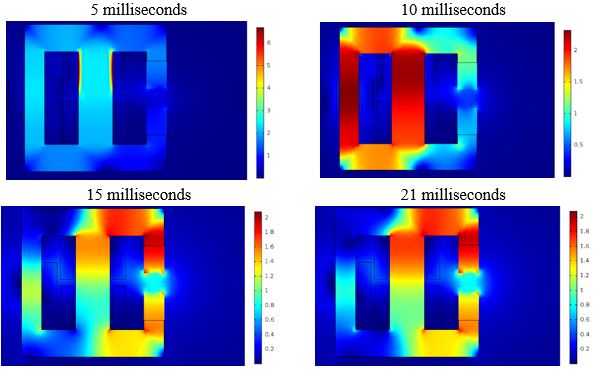}
\caption{Flux density distribution for Model A during fault mode}
\label{fig:Fig. 2.26}
\end{figure}

We now examine the flux density distribution at different points in these three models and make a comparison between them. First, we show how the flux density changes in Model A, which is in the form of Fig.~\ref{fig:Fig. 2.25} and Fig.~\ref{fig:Fig. 2.26} (for an air gap length of 0.3 m). In Fig.~\ref{fig:Fig. 2.25}, there is a flux density reduction in middle leg that increases near the fault point. After fault occurrence, there is a huge drop of flux density that density’s concentration is on the top of middle leg at 5 ms. Then, the flux density distributes to other legs after 5 ms that middle leg experiences the highest flux density. Afterwards, the high-density transfer to the right leg.

As you can see, the flux density gradually decreases in the central leg and goes out of saturation until it rises again in about 21 ms, but it has never entered deep saturation in this case. Now, the flux density is very high in some positions, due to the fault originating the passage of current through the ac windings. The middle leg is deeply saturated, which is done using the dc auxiliary winding in 10 ms. Flux density is gradually directed from the middle leg to the leg at an air gap, creating little saturation in those areas. Also, the flux density distribution is given for these three models as an example in 5 ms that the diagram has already been given.

\begin{figure}
\centering
\includegraphics[height=8cm]{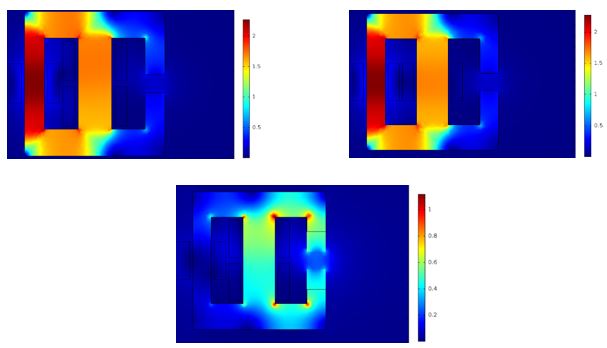}
\caption{Comparison of flux density distribution at 5 milliseconds}
\label{fig:Fig. 2.27}
\end{figure}

\section{Inductive SCFCL}
\begin{figure}
\centering
\includegraphics[height=5.5cm]{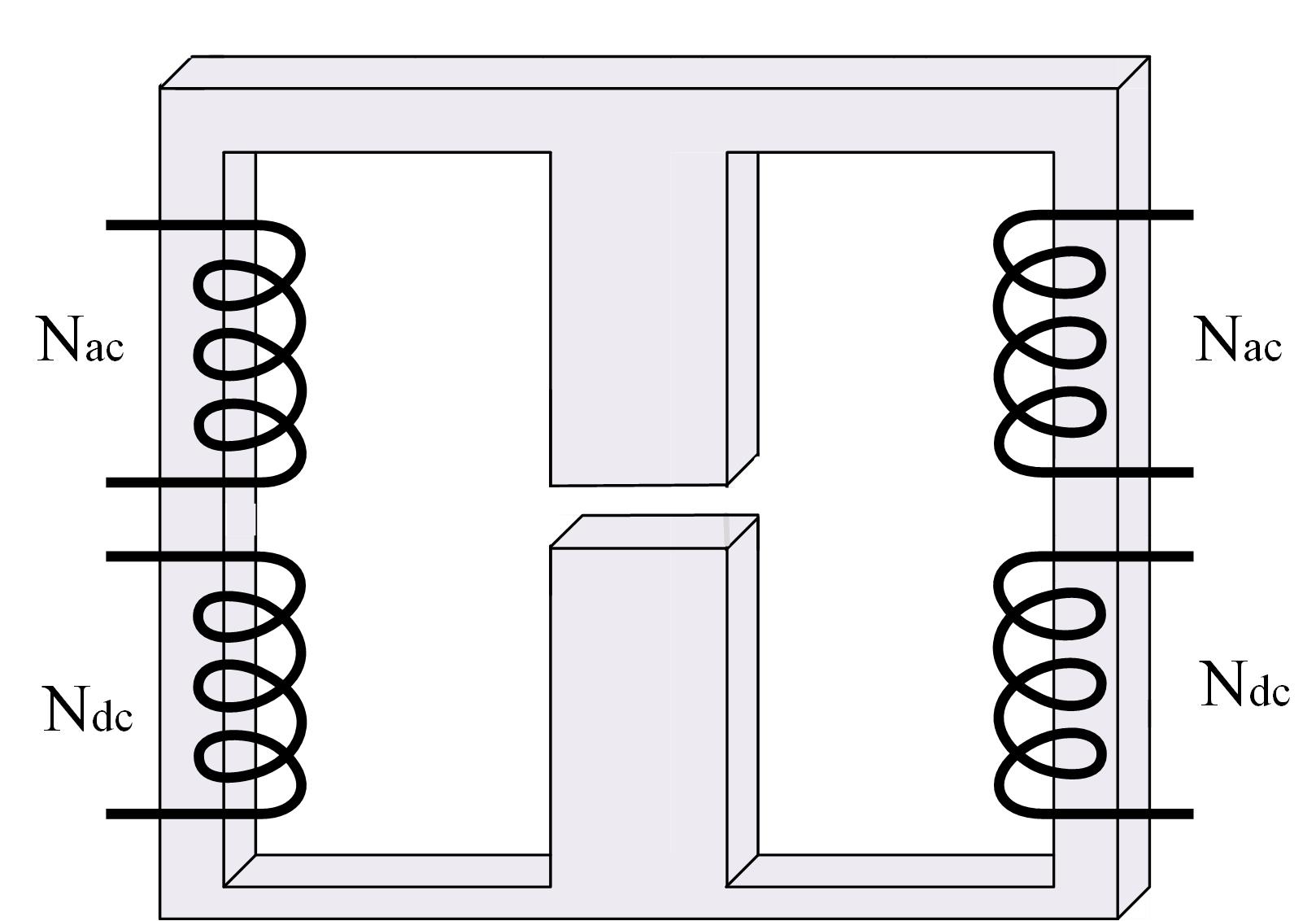}
\caption{Single phase inductive SCFCL}
\label{fig:Fig. 2.28}
\end{figure}

The configuration of a single phase inductive SCFCL is shown in Fig.~\ref{fig:Fig. 2.28}. In this structure, they alternate from one half-cycle to the other half-cycle because the two legs have an ac winding separately, and the leg with the air gap directs the ac flux produced in this leg. Therefore, there are three magnetic legs in this architecture, each with dc and ac coils. The ac coils are placed to the transmission line in series, whilst the dc coils are connected to the direct current source in series \cite{cvoric20163d,cvoric2008comparison,pirhadi2020design}.

The dc windings are connected in a way that the dc flux receives a rotational current. As a result, the outside legs of the core are saturated, but the dc flux does not pass through the central leg. So, dc saturation level does not rely on the length of the air gap ($l_{gap}$), so we have \cite{linden2019design}:

\begin{equation}\label{13}
 B_{mid}=0, \quad B_{o} = \mu_{r,sat} \times \frac{N_{dc} I_{dc}}{l_{mean,outer}} + (B_{sat} - \mu_{sat} H_{sat})
\end{equation}

Where $B_{o}$ and $B_{mid}$ and are the dc flux densities in the outer and central legs, correspondingly, and $l_{mean,outer}$ is the average lengths of the outer leg and $H_{sat}$ is the saturation values of the magnetic field. The central leg makes a parallel route for the ac flux, allowing the air gap to be merely in the ac magnetic circuit. After the fault commences, the left and right legs of the core alternate from saturation mode. The connection of the ac windings is during a half cycle, the ac and dc fluxes are in the same direction but in the other outer leg are in opposite directions. Since the reluctance of the right leg is quite high, the ac flux of the left leg blocks its pathway via the center part due to saturation. Inductance of SCFCL can be expressed during normal and fault conditions as Eqs. (\ref{14}) and (\ref{15}) \cite{commins2012three,commins2012analytical,zhou2020inductive,5415715,pellecchia2016development}:

\begin{equation}\label{14}
 L_{FCL,sat} = \mu_{0} \times \frac{N^2_{ac} \times A_{core,outer}}{l_{mean,outer}+l_{gap}} = \mu_{0} \times \frac{N^2_{ac} \times A_{core,outer}}{({l_{mean,outer}}/{l_{gap}}+1) \times l_{gap}}
\end{equation}

\begin{equation}\label{15}
 L_{FCL,lin} = \mu_{0} \times \frac{N^2_{ac} \times 2A_{core,outer}}{l_{gap}} = f(2A_{core,outer})
\end{equation}

Where $L_{FCL,sat}$ is FCL inductance in saturation mode, $A_{core}$, outer cross-sectional area of the outer leg, and $L_{FCL,lin}$ is FCL inductance out of saturated mode (in fault mode).

\section{SCFCL with DC Single-Core Short-Circuited Windings}
\begin{figure}
\centering
\includegraphics[height=5.5cm]{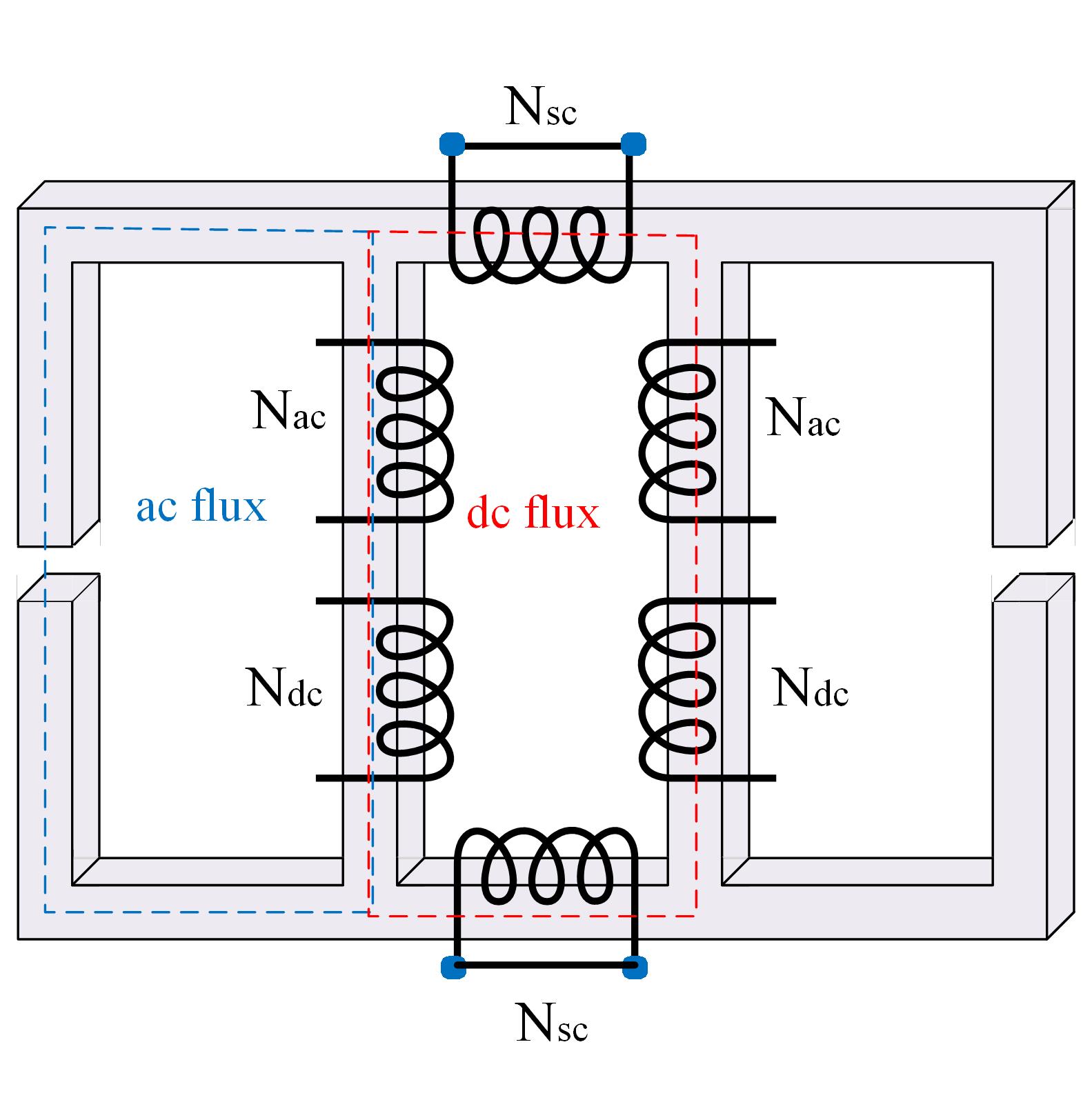}
\caption{Single-core SCFCL configuration with short-circuited windings}
\label{fig:Fig. 2.29}
\end{figure}

SCFCL with dc short-circuited windings uses two cores in each phase. The two cores can be combined to further reduce the magnetic materials. This topology is illustrated in Fig.~\ref{fig:Fig. 2.29}. As you can observe, because the dc leg's main purpose was to make a low reluctance path for the dc flux, it was eliminated. In this setup, an extra ac leg takes over the dc leg's role. The ac flux is directed from each ac winding to the air gap legs using short-circuited windings \cite{baferani2018novel,commins2012analytical,pellecchia2016development,vilhena2018design}.

\begin{figure}
\centering
\includegraphics[height=4.5cm]{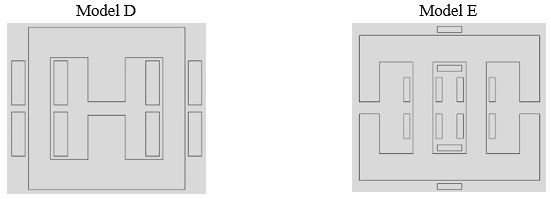}
\caption{SCFCL configurations including an inductive saturated core FCL (Model D) and a single-core saturated core FCL with short-circuited windings (Model E)}
\label{fig:Fig. 2.30}
\end{figure}

\begin{figure}
\centering
\includegraphics[height=6cm]{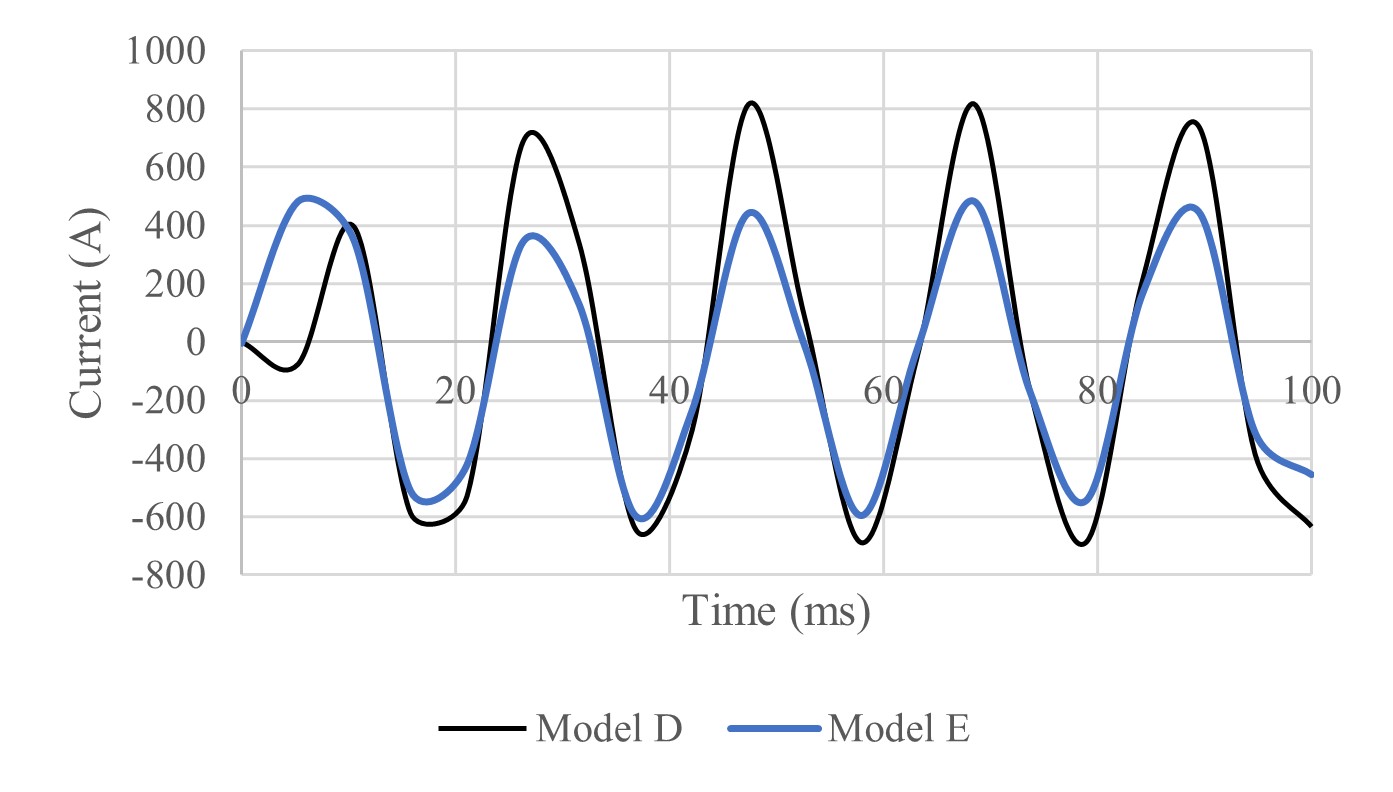}
\caption{Current passing through the system with SCFCL in a normal mode}
\label{fig:Fig. 2.31}
\end{figure}

In this section, we evaluate the simulation results of models D and E, which we define as the models in Fig.~\ref{fig:Fig. 2.30}. The current passing through the system in the presence of FCL in normal operating conditions has been investigated in this case. By mentioning to Fig.~\ref{fig:Fig. 2.31}, it can be seen that these two models have larger impedances than models A, B and C in the normal operating mode which is not desirable and their current has reached up to 400 A or more even in the first peak, although the rated current of the system has a peak of 1000 A. Model D is more desirable than model E in this state (because model E has a larger number of windings and therefore it has more resistance and impedance in a normal operation). There is a drop at the outset of electric current of Model D because of air gap presence at the middle leg.

\begin{figure}
\centering
\includegraphics[height=6cm]{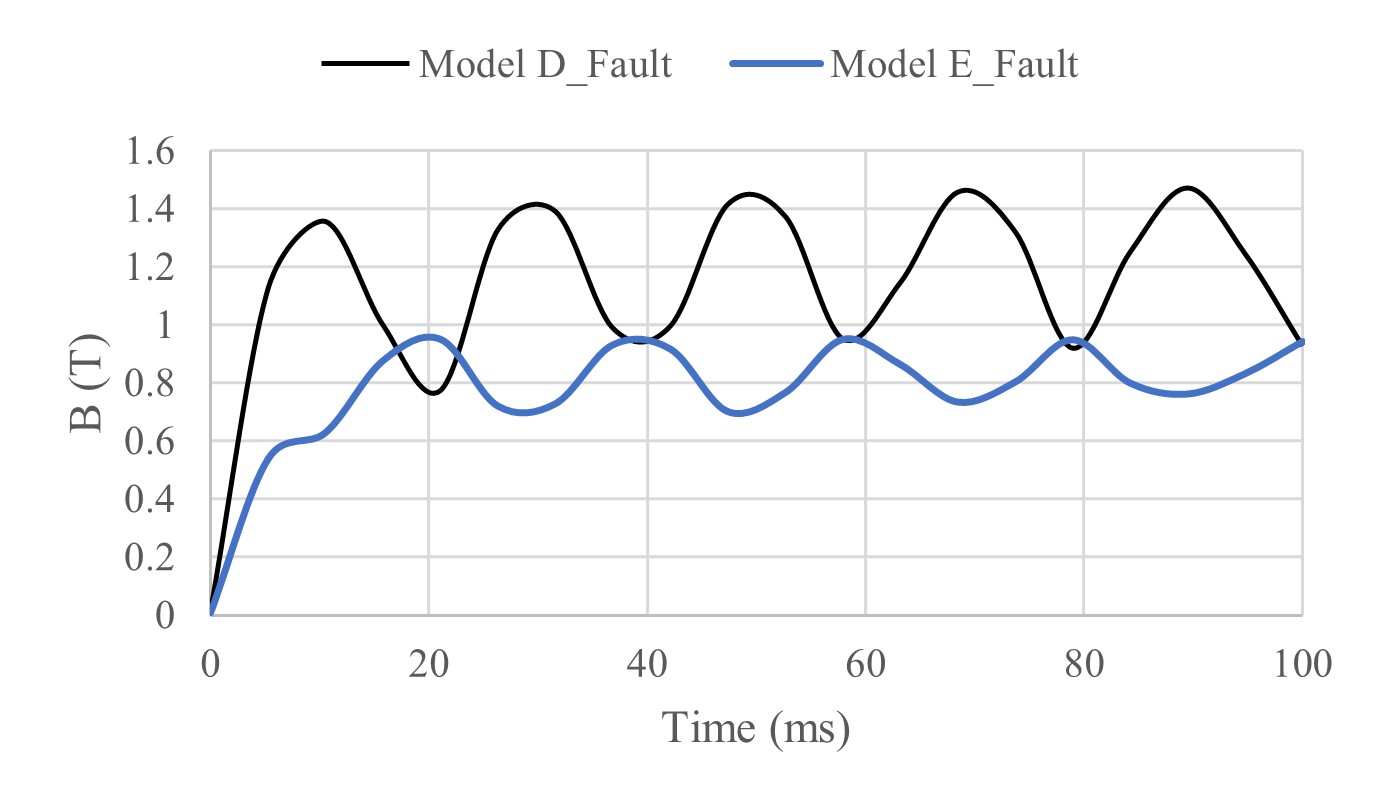}
\caption{Flux density of iron core during fault mode for Models D and E}
\label{fig:Fig. 2.32}
\end{figure}

Model D has better saturation than model E, because the outer legs in model D have dc winding and are better saturated, while on model E, the dc winding has short-circuited. It causes the iron core to saturate around it later, as shown in Fig.~\ref{fig:Fig. 2.32} that the flux density changes at fault mode for models D and E.

\begin{figure}
\centering
\includegraphics[height=6cm]{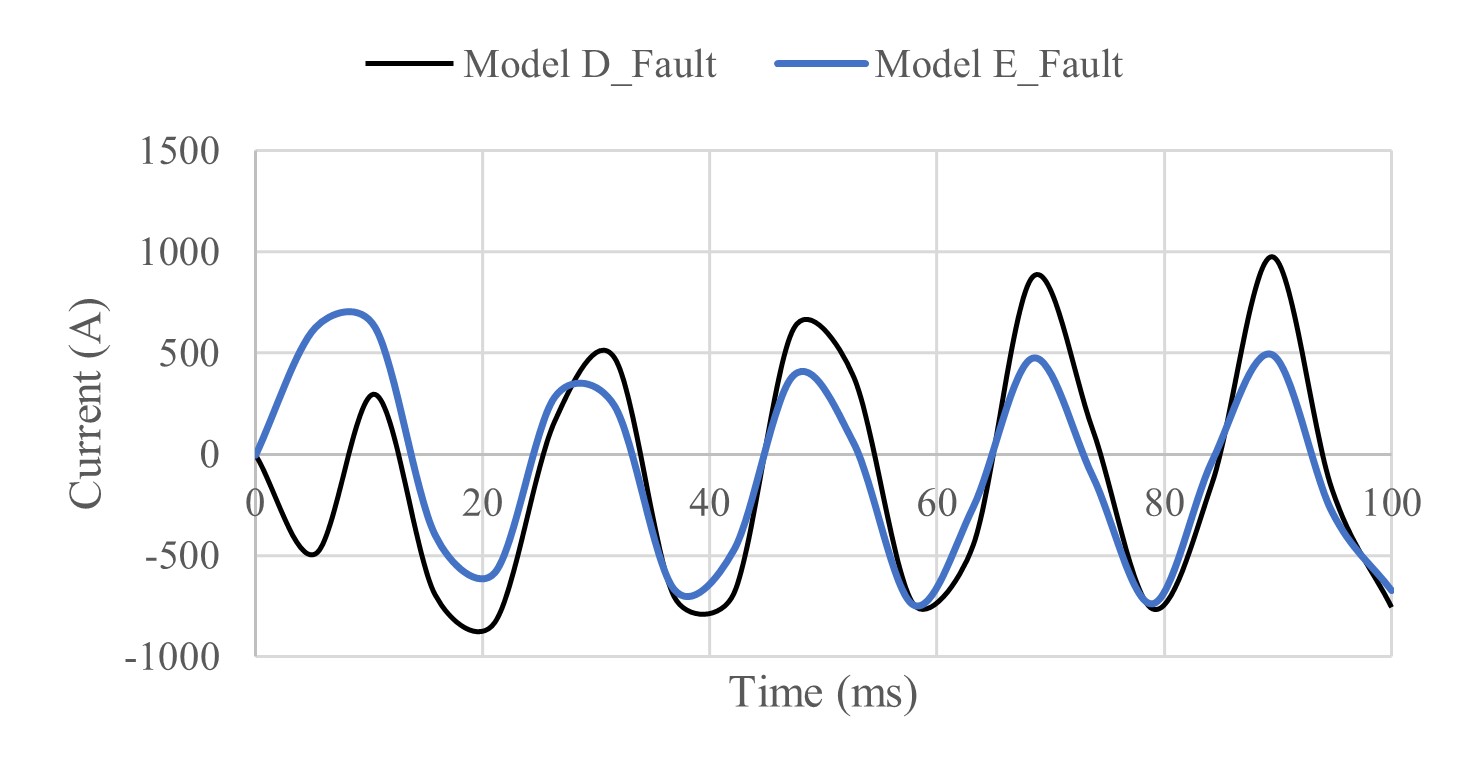}
\caption{Current passing through during fault condition with SCFCL}
\label{fig:Fig. 2.33}
\end{figure}

It is also possible to show the amount of current in fault mode for models D and E, which due to the further voltage drops of them in which their limitation is even less than the nominal current (Fig.~\ref{fig:Fig. 2.33}). Fig.~\ref{fig:Fig. 2.33} observes higher current amplitude than Fig.~\ref{fig:Fig. 2.31}. These structures have slight effect on reducing the fault in comparison with configurations introduced in previous section and has more material consuming.

\begin{figure}
\centering
\includegraphics[height=8cm]{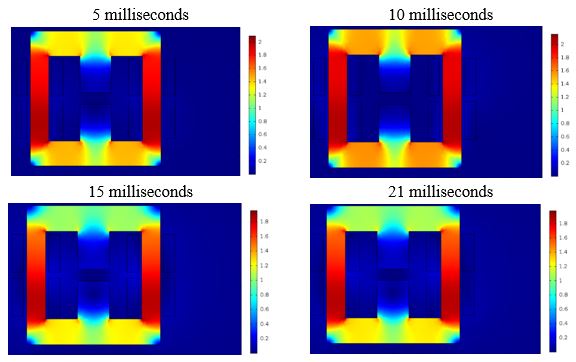}
\caption{Flux density distribution for Model D during normal mode}
\label{fig:Fig. 2.34}
\end{figure}

\begin{figure}
\centering
\includegraphics[height=8cm]{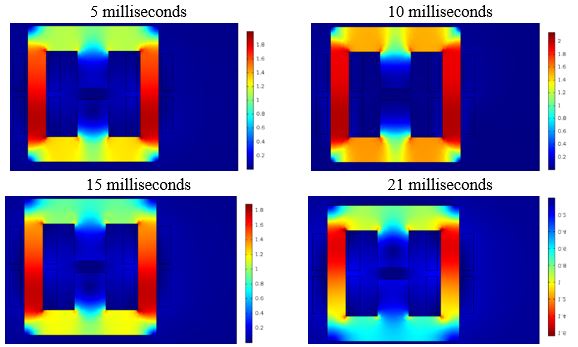}
\caption{Flux density distribution for Model D during fault mode}
\label{fig:Fig. 2.35}
\end{figure}

Now, we want to evaluate the flux density distribution in Model D, which are in Fig.~\ref{fig:Fig. 2.34} and Fig.~\ref{fig:Fig. 2.35} for different times in normal and fault states. By comparing the above two states of flux density distribution, because of the occurrence of a fault, two outer legs gradually deviate from the saturation state, so the flux density is greater than the fault state in the normal mode. It should also be not-ed that we must apply much dc current to dc winding to fully enter the saturation zone at the beginning of the design that the analogical diagram for two normal and fault modes for model D is shown for 100 milliseconds.

\begin{figure}
\centering
\includegraphics[height=6cm]{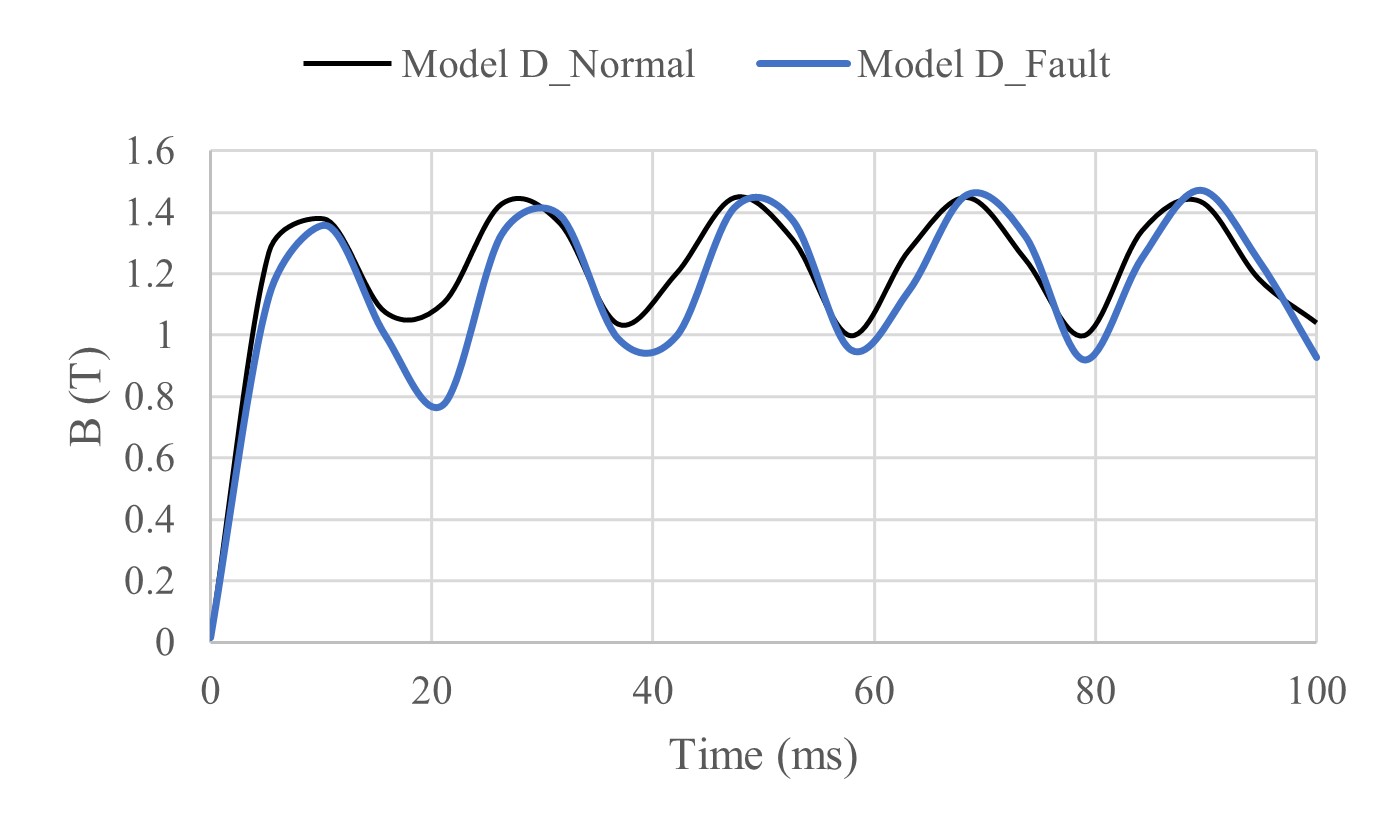}
\caption{Flux density for Model D during normal and fault operations}
\label{fig:Fig. 2.36}
\end{figure}

\begin{figure}
\centering
\includegraphics[height=8cm]{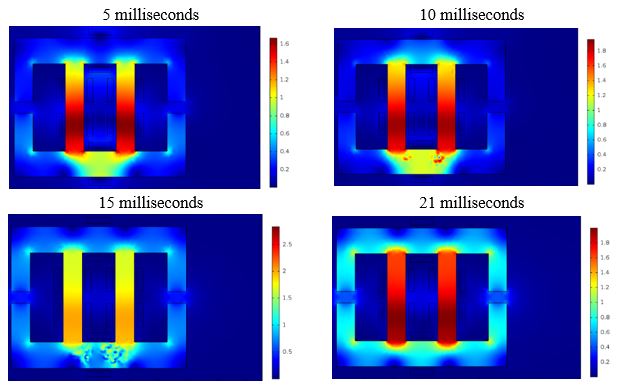}
\caption{Flux density distribution for Model E during normal operation}
\label{fig:Fig. 2.37}
\end{figure}
It is also possible to provide a flux density distribution for Model E, which is as follows. As it can be perceived, the middle legs are not deeply saturated from the beginning, because of the presence of short-circuit windings at the top and bottom of the core, and a large dc current must be applied to the winding to achieve this. The middle legs become gradually saturated over time and the flux slowly disperses at the surface of the iron core, with the middle legs alternating between saturated and unsaturated modes. There is a summary of proposed configurations and their specifications as Table \ref{table 2.3}.

\begin{center}
\begin{table}
    %\centering
\caption{Summary of proposed SCFCL configurations}

\label{table 2.3}
    
     \setlength{\leftmargini}{1cm}
    \begin{tabular}{| m{7cm} | m{6cm} |}
        \hline
     \centering \textbf{SCFCL Configuration} & \textbf{Design Specifications} \\
        \hline
       \centering \includegraphics[height=2.5cm]{figures/Fig.2.11.jpg} & 
        \begin{itemize} 
            \item Every phase has two cores. 
            \item Partial magnetic separation of dc \& ac fluxes
            \item An air gap in side leg
            \item Voltage is induced in dc winding
            \item FCL impedance is declined during normal conditions
        \end{itemize} \\ 
        \hline
        \centering \includegraphics[height=2cm]{figures/Fig.2.9.jpg} & 
        \begin{itemize} 
            \item Every phase has two cores 
            \item There is no voltage induced
            \item 100\% magnetic separation of dc \& ac fluxes
            \item An air gap in sidelong leg
            \item There is no deep saturation in the middle leg because of one ac winding
            \item FCL impedance is increased during normal conditions
        \end{itemize} \\ 
        \hline
        \centering \includegraphics[height=2.5cm]{figures/Fig.2.13.jpg} & 
        \begin{itemize} 
            \item Every phase has two cores 
            \item Partial magnetic separation of dc \& ac fluxes
            \item An air gap in sidelong leg
            \item Volume of winding material is mitigated because of short-circuited dc winding
        \end{itemize} \\ 
        \hline
        \centering \includegraphics[height=2.5cm]{figures/Fig.2.28.jpg} & 
        \begin{itemize} 
            \item Existence of an air gap in middle leg 
            \item Partial magnetic separation of dc \& ac fluxes
            \item Single phase inductive SCFCL (every phase has a core)
            \item Conduction of ac flux across the leg with air gap is sufficient
        \end{itemize} \\ 
        \hline
        \centering \includegraphics[height=3cm]{figures/Fig.2.29.jpg} & 
        \begin{itemize} 
            \item Partial magnetic separation of dc \& ac fluxes 
            \item Single phase inductive SCFCL (every phase has a core)
            \item Changes in flux distribution because of winding positions
        \end{itemize} \\ 
        \hline
    \end{tabular}
\end{table}
\end{center}

\section{Conclusion}
Nowadays fault current limiters can have considerable effect on any failure or outage in power systems. They are very beneficial to postpone the replacement of power devices and take an important role in mitigation of fault impacts. Saturated-core FCLs can be attractive because of saturation features and prompt changes of flux between the different parts of an iron core. Suitable designs of this type of FCLs in terms of material, volume, type of winding placement, etc. can be helpful for industrial engineers to utilize an efficient configuration regarding their applicability. This device can remarkably reduce the vulnerability of a power grid. This chapter presents an analysis of different configurations of SCFCLs that their design is relied on number of windings, winding positions, etc. Various parameters of this protective device including electric current, magnetic flux density, magnetic field strength, etc. are assessed and compared. Numerical analysis with finite element method is carried out to provide a comprehensive information related to SCFCLs to demonstrate the variables’ changes during normal operation or fault mode better.

\bibliography{fcl_ref}

\begin{thebibliography}{}

\bibitem[Alam et~al., 2018]{alam2018fault}
Alam, M.~S., Abido, M. A.~Y., and El-Amin, I. (2018).
\newblock Fault current limiters in power systems: A comprehensive review.
\newblock {\em Energies}, 11(5):1025.

\bibitem[Badakhshan et~al., 2018]{badakhshan2018flux}
Badakhshan, M. et~al. (2018).
\newblock Flux-lock type of superconducting fault current limiters: A
  comprehensive review.
\newblock {\em Physica C: Superconductivity and its Applications}, 547:51--54.

\bibitem[Baferani et~al., 2018]{baferani2018novel}
Baferani, M.~A., Chalaki, M.~R., Fahimi, N., Shayegani, A.~A., and Niayesh, K.
  (2018).
\newblock A novel arrangement for improving three phase saturated-core fault
  current limiter (scfcl).
\newblock In {\em 2018 IEEE Texas Power and Energy Conference (TPEC)}, pages
  1--6. IEEE.

\bibitem[Baimel et~al., 2021]{baimel2021new}
Baimel, D., Chowdhury, N., Belikov, J., and Levron, Y. (2021).
\newblock New type of bridge fault current limiter with reduced power losses
  for transient stability improvement of dfig wind farm.
\newblock {\em Electric Power Systems Research}, 197:107293.

\bibitem[Branco et~al., 2010]{branco2010proposal}
Branco, P.~C., Almeida, M., and Dente, J. (2010).
\newblock Proposal for an rms thermoelectric model for a resistive-type
  superconducting fault current limiter (sfcl).
\newblock {\em Electric power systems research}, 80(10):1229--1239.

\bibitem[Chen et~al., 2016]{chen2016parameter}
Chen, B., Wei, L., Tian, C., Lei, Y., and Yuan, J. (2016).
\newblock Parameter design and performance investigation of a novel bridge-type
  saturated core fault current limiter.
\newblock {\em IEEE Transactions on Power Delivery}, 32(2):1049--1057.

\bibitem[Commins and Moscrop, 2012a]{commins2012analytical}
Commins, P.~A. and Moscrop, J.~W. (2012a).
\newblock Analytical nonlinear reluctance model of a single-phase saturated
  core fault current limiter.
\newblock {\em IEEE Transactions on Power Delivery}, 28(1):450--457.

\bibitem[Commins and Moscrop, 2012b]{commins2012three}
Commins, P.~A. and Moscrop, J.~W. (2012b).
\newblock Three phase saturated core fault current limiter performance with a
  floating neutral.
\newblock In {\em 2012 IEEE Electrical Power and Energy Conference}, pages
  249--254. IEEE.

\bibitem[Cvoric et~al., 2009a]{cvoric2009new}
Cvoric, D., De~Haan, S., and Ferreira, J. (2009a).
\newblock New saturable-core fault current limiter topology with reduced core
  size.
\newblock In {\em 2009 IEEE 6th International Power Electronics and Motion
  Control Conference}, pages 920--926. IEEE.

\bibitem[Cvoric et~al., 2008]{cvoric2008comparison}
Cvoric, D., de~Haan, S.~W., and Ferreira, J. (2008).
\newblock Comparison of the four configurations of the inductive fault current
  limiter.
\newblock In {\em 2008 IEEE Power Electronics Specialists Conference}, pages
  3967--3973. IEEE.

\bibitem[Cvoric et~al., 2009b]{5415715}
Cvoric, D., de~Haan, S. W.~H., and Ferreira, J.~A. (2009b).
\newblock Guidelines for 2d/3d fe transient modeling of inductive
  saturable-core fault current limiters.
\newblock In {\em 2009 International Conference on Electric Power and Energy
  Conversion Systems, (EPECS)}, pages 1--6.

\bibitem[Cvoric et~al., 2016]{cvoric20163d}
Cvoric, D., Lahaye, D., de~Haan, S., and Ferreira, B. (2016).
\newblock 3d nonlinear transient field-circuit modeling of inductive fault
  current limiters.
\newblock {\em International Journal of Numerical Modelling: Electronic
  Networks, Devices and Fields}, 29(2):354--360.

\bibitem[Gunawardana et~al., 2016]{gunawardana2016transient}
Gunawardana, S., Commins, P., Moscrop, J., and Perera, S. (2016).
\newblock Transient modeling of saturated core fault current limiters.
\newblock {\em IEEE Transactions on Power Delivery}, 31(5):2008--2017.

\bibitem[Han et~al., 2018]{han2018fault}
Han, T.-H., Ko, S.-C., and Lim, S.-H. (2018).
\newblock Fault current limiting characteristics of transformer-type
  superconducting fault current limiter due to winding direction of additional
  circuit.
\newblock {\em IEEE Transactions on Applied Superconductivity}, 28(3):1--6.

\bibitem[Jia et~al., 2017]{jia2017numerical}
Jia, Y., Ainslie, M.~D., Hu, D., and Yuan, J. (2017).
\newblock Numerical simulation and analysis of a saturated-core-type
  superconducting fault current limiter.
\newblock {\em IEEE Transactions on Applied Superconductivity}, 27(4):1--5.

\bibitem[Kim et~al., 2010]{kim2010study}
Kim, J.-S., Lim, S.-H., and Kim, J. (2010).
\newblock Study on protection coordination of a flux-lock type sfcl with
  over-current relay.
\newblock {\em IEEE Transactions on Applied Superconductivity},
  20(3):1159--1163.

\bibitem[Li et~al., 2019]{li2019current}
Li, B., Cui, H., Jing, F., Li, B., and Liu, Y. (2019).
\newblock Current-limiting characteristics of saturated iron-core fault current
  limiters in vsc-hvdc systems based on electromagnetic energy conversion
  mechanism.
\newblock {\em Journal of Modern Power Systems and Clean Energy},
  7(2):412--421.

\bibitem[Lim, 2011]{lim2011analysis}
Lim, S. (2011).
\newblock Analysis on current limiting characteristics of a flux-lock type sfcl
  with two triggering current levels.
\newblock {\em Physica C: Superconductivity and its Applications},
  471(21-22):1354--1357.

\bibitem[Linden et~al., 2019]{linden2019design}
Linden, J., Nikulshin, Y., Friedman, A., Yeshurun, Y., and Wolfus, S. (2019).
\newblock Design optimization of a permanent-magnet saturated-core
  fault-current limiter.
\newblock {\em Energies}, 12(10):1823.

\bibitem[Linden et~al., 2020]{linden2020phase}
Linden, J., Nikulshin, Y., Friedman, A., Yeshurun, Y., and Wolfus, S. (2020).
\newblock Phase-coupling effects in three-phase inductive fault-current limiter
  based on permanent magnets.
\newblock {\em IEEE Transactions on Magnetics}, 56(2):1--7.

\bibitem[Liu et~al., 2021]{liu2021design}
Liu, Y., Guan, L., Tan, Z., Yang, K., Guo, F., Chen, Y., Liu, R., and Zheng, F.
  (2021).
\newblock Design of a multifunction novel flexible fault current limiter for ac
  distribution network.
\newblock {\em Plos one}, 16(4):e0245956.

\bibitem[Naphade et~al., 2021]{naphade2021experimental}
Naphade, V., Ghate, V., and Dhole, G. (2021).
\newblock Experimental analysis of saturated core fault current limiter
  performance at different fault inception angles with varying dc bias.
\newblock {\em International Journal of Electrical Power \& Energy Systems},
  130:106943.

\bibitem[Nikulshin et~al., 2016]{nikulshin2016saturated}
Nikulshin, Y., Wolfus, Y., Friedman, A., Yeshurun, Y., Rozenshtein, V.,
  Landwer, D., and Garbi, U. (2016).
\newblock Saturated core fault current limiters in a live grid.
\newblock {\em IEEE Transactions on Applied Superconductivity}, 26(3):1--4.

\bibitem[Ouali and Cherkaoui, 2020]{ouali2020integration}
Ouali, S. and Cherkaoui, A. (2020).
\newblock Integration of fault current limiters in electric power systems: A
  review.
\newblock {\em IEEE Systems Journal}, 15(3):4470--4479.

\bibitem[Pellecchia et~al., 2016]{pellecchia2016development}
Pellecchia, A., Klaus, D., Masullo, G., Marabotto, R., Morandi, A., Fabbri, M.,
  Goodhand, C., and Helm, J. (2016).
\newblock Development of a saturated core fault current limiter with open
  magnetic cores and magnesium diboride saturating coils.
\newblock {\em IEEE Transactions on Applied Superconductivity}, 27(4):1--7.

\bibitem[Pirhadi et~al., 2020]{pirhadi2020design}
Pirhadi, A., Shayan, H., and Bina, M.~T. (2020).
\newblock Design of an unsaturated core-based fault current limiter to tackle
  unsymmetrical faults.
\newblock {\em Electric Power Systems Research}, 187:106482.

\bibitem[Ruiz et~al., 2014]{ruiz2014resistive}
Ruiz, H.~S., Zhang, X., and Coombs, T. (2014).
\newblock Resistive-type superconducting fault current limiters: concepts,
  materials, and numerical modeling.
\newblock {\em IEEE Transactions on Applied Superconductivity}, 25(3):1--5.

\bibitem[Safaei et~al., 2020]{safaei2020survey}
Safaei, A., Zolfaghari, M., Gilvanejad, M., and Gharehpetian, G.~B. (2020).
\newblock A survey on fault current limiters: Development and technical
  aspects.
\newblock {\em International Journal of Electrical Power \& Energy Systems},
  118:105729.

\bibitem[Shen et~al., 2018a]{shen2018three}
Shen, H., Mei, F., Zheng, J., Sha, H., and She, C. (2018a).
\newblock Three-phase saturated-core fault current limiter.
\newblock {\em Energies}, 11(12):3471.

\bibitem[Shen et~al., 2018b]{shen2018study}
Shen, H., Zheng, J., Sha, H., She, C., and Mei, F. (2018b).
\newblock Study of a new structure of three-phase saturated-core fault current
  limiter.
\newblock In {\em 2018 International Conference on Smart Grid and Clean Energy
  Technologies (ICSGCE)}, pages 239--243. IEEE.

\bibitem[Tan et~al., 2015]{tan2015resistive}
Tan, Y.~X., Yang, K., Xiang, B., Yan, J., San~Geng, Y., Liu, Z.~Y., Wang,
  J.~H., and Yanabu, S. (2015).
\newblock Resistive type superconducting fault current limiter and current
  flowing time.
\newblock In {\em 2015 IEEE International Conference on Applied
  Superconductivity and Electromagnetic Devices (ASEMD)}, pages 290--291. IEEE.

\bibitem[Tripathi and Chatterjee, 2021]{tripathi2021real}
Tripathi, P.~M. and Chatterjee, K. (2021).
\newblock Real-time implementation of ring based saturated core fault current
  limiter to improve fault ride through capability of dfig system.
\newblock {\em International Journal of Electrical Power \& Energy Systems},
  131:107040.

\bibitem[Vilhena et~al., 2018]{vilhena2018design}
Vilhena, N., Murta-Pina, J., Pronto, A., and {\'A}lvarez, A. (2018).
\newblock A design methodology for the optimization of three-phase sfcl of
  saturated cores type.
\newblock {\em IEEE Transactions on Applied Superconductivity}, 28(4):1--5.

\bibitem[Wei et~al., 2018]{wei2018performance}
Wei, L., Chen, B., Liu, Y., Tian, C., Yuan, J., Bu, Y., and Zhu, T. (2018).
\newblock Performance investigation and optimization of a novel hybrid
  saturated-core fault-current limiter considering the leakage effect.
\newblock {\em Energies}, 11(1):61.

\bibitem[Wei et~al., 2020]{wei2020limiting}
Wei, L., Chen, B., and Yuan, J. (2020).
\newblock Limiting mechanism and equivalent analytical model of a bridge-type
  saturated core fault current limiter.
\newblock {\em Electric Power Systems Research}, 180:106134.

\bibitem[Yamaguchi et~al., 2004]{yamaguchi2004characteristics}
Yamaguchi, H., Kataoka, T., Yaguchi, K., Fujita, S., Yoshikawa, K., and Kaiho,
  K. (2004).
\newblock Characteristics analysis of transformer type superconducting fault
  current limiter.
\newblock {\em IEEE transactions on applied superconductivity}, 14(2):815--818.

\bibitem[Yuan et~al., 2020a]{yuan2020saturated}
Yuan, J., Gan, P., Zhang, Z., Zhou, H., Wei, L., and Muramatsu, K. (2020a).
\newblock Saturated-core fault current limiters for ac power systems: Towards
  reliable, economical and better performance application.
\newblock {\em High Voltage}, 5(4):416--424.

\bibitem[Yuan et~al., 2020b]{yuan2020optimized}
Yuan, J., Zhang, Z., Zhou, H., Gan, P., and Chen, H. (2020b).
\newblock Optimized design method of permanent magnets saturated core fault
  current limiters for hvdc applications.
\newblock {\em IEEE Transactions on Power Delivery}, 36(2):721--730.

\bibitem[Yuan et~al., 2018]{yuan2018novel}
Yuan, J., Zhou, H., Gan, P., Zhong, Y., Gao, Y., Muramatsu, K., Du, Z., and
  Chen, B. (2018).
\newblock A novel concept of fault current limiter based on saturable core in
  high voltage dc transmission system.
\newblock {\em Aip Advances}, 8(5):056636.

\bibitem[Zhang and Zhang, 2019]{zhang2019viable}
Zhang, X. and Zhang, Y. (2019).
\newblock A viable approach for limiting fault currents in electric networks.
\newblock {\em IEEJ Transactions on Electrical and Electronic Engineering},
  14(4):556--560.

\bibitem[Zhao et~al., 2014]{zhao2014performance}
Zhao, Y., Krause, O., Saha, T.~K., and Li, Y. (2014).
\newblock Performance analysis of flux-lock type sfcl influenced by
  characteristics of two coils.
\newblock In {\em 2014 IEEE PES General Meeting| Conference \& Exposition},
  pages 1--5. IEEE.

\bibitem[Zhou et~al., 2020a]{zhou2020hybrid}
Zhou, H., Yuan, J., and Chen, F. (2020a).
\newblock Hybrid-material based saturated core fcl in hvdc system: Modeling,
  analyzing and performance testing.
\newblock {\em IEEE Transactions on Industrial Electronics},
  68(12):11858--11869.

\bibitem[Zhou et~al., 2020b]{zhou2020inductive}
Zhou, H., Yuan, J., Chen, F., and Chen, B. (2020b).
\newblock Inductive fault current limiters in vsc-hvdc systems: A review.
\newblock {\em IEEE Access}, 8:38185--38197.

\bibitem[Zhou et~al., 2019]{zhou2019performance}
Zhou, H., Yuan, J., Chen, F., Chen, B., and Muramatsu, K. (2019).
\newblock Performance investigation on a novel high inductance changing ratio
  mmc-based direct current system saturated core fcl.
\newblock {\em IEEE Transactions on Power Delivery}, 35(3):1502--1514.

\end{thebibliography}

\end{document}